\documentclass{article}
\usepackage[utf8]{inputenc}
\usepackage{amsmath}
\usepackage{mathtools}
\usepackage{breqn}
\usepackage{amsfonts}
\usepackage{enumerate}
\usepackage{amssymb}
\usepackage{comment}
\usepackage{epsfig}
\usepackage{graphicx,url}
\usepackage{bm}
\usepackage{empheq}
\usepackage[top=2.5cm,bottom=2.5cm,right=2cm,left=2cm]{geometry}
\usepackage{float} 
\usepackage{amsthm}
\usepackage{bigints}
\usepackage{stackengine}
\usepackage{color} 
\usepackage{bbm}
\usepackage{appendix}
\usepackage[nottoc]{tocbibind}

\stackMath

\usepackage{enumerate}

\usepackage{epstopdf}

\usepackage{zref-base}
\usepackage{atveryend}
\makeatletter
\AfterLastShipout{%
	\if@filesw   
	\begingroup
	\zref@setcurrent{default}{\the\value{equation}}%
	\zref@wrapper@immediate{%
		\zref@labelbyprops{eq-last}{default}%
	}%
	\endgroup
	\fi
}
\makeatother

\DeclareMathOperator{\map}{map}

\DeclareMathOperator{\ind}{ind}

\DeclareMathOperator{\nmap}{n_{\map}}
\DeclareMathOperator{\nind}{n_{\ind}}

\DeclareMathOperator{\free}{id}

\DeclareMathOperator{\car}{car}

\newcommand{\mb}{\mathbf}

\newcommand{\Le}{\left}
\newcommand{\Ri}{\right}
\newcommand{\lla}{\left \langle}
\newcommand{\rra}{\right \rangle}
\newcommand{\p}{\partial}
\newcommand{\f}{\frac}
\newcommand{\mc}{\mathcal}
\newcommand{\mr}{\mathrm}
\newcommand{\bs}{\boldsymbol}
 
\newcommand{\nn}{\nonumber} 
\newcommand{\llv}{\left\lvert}
\newcommand{\rrv}{\right\rvert}
\newcommand{\lv}{\lvert}
\newcommand{\rv}{\rvert}  
 
\newcommand{\lrar}{\leftrightarrow}

\DeclarePairedDelimiter\ceil{\lceil}{\rceil}
\DeclarePairedDelimiter\floor{\lfloor}{\rfloor}
\def\multiset#1#2{\ensuremath{\left(\kern-.3em\left(\genfrac{}{}{0pt}{}{#1}{#2}\right)\kern-.3em\right)}}

\begin{document}
	
\begin{center}
	\Large
	\textbf{Memory effects in the Fermi-Pasta-Ulam Model}\\ 
	
	\hspace{10pt}
	
	\large
	Graziano Amati$^1$, Hugues Meyer$^{1,2}$, Tanja Schilling$^1$ \\
	
	\hspace{10pt}

	\small
	$^1$ Physikalisches Institut, Albert-Ludwigs-Universität, 79104 Freiburg, Germany \\
	$^2$ Unit in Engineering Science, Université du Luxembourg, L-4364 Esch-sur-Alzette, Luxembourg
\end{center}

\hspace{10pt}

\section*{Abstract}
We study the Intermediate Scattering Function (ISF) of the strongly-nonlinear Fermi-Pasta Ulam Model at thermal equilibrium, using both numerical and analytical methods. From the molecular dynamics simulations we distinguish two limit regimes, as the system behaves as an ideal gas at high temperature and as a harmonic chain for low excitations. At intermediate temperatures the ISF relaxes to equilibrium in a nontrivial fashion. We then calculate analytically the Taylor coefficients of the ISF to arbitrarily high orders (the specific, simple shape of the two-body interaction allows us to derive an iterative scheme for these.) The results of the recursion are in good agreement with the numerical ones. Via an estimate of the complete series expansion of the scattering function, we can reconstruct within a certain temperature range its coarse-grained dynamics. This is governed by a memory-dependent Generalized Langevin Equation (GLE), which can be derived via projection operator techniques. Moreover, by analyzing the first series coefficients of the ISF, we can extract a parameter associated to the strength of the memory effects in the dynamics.
\section{Introduction}
The Fermi-Pasta-Ulam Model (FPU) consists of a one-dimensional chain of $N$ particles interacting through a nearest-neighbor potential. The system was at first conceived as a simple numerical tool to probe the spread of chaos in nonlinear dynamical systems. It is typically interpreted as a nonlinear crystal, while at the same time it can map a broad class of other realistic systems such as DNA structures \cite{dna} and polymer chains \cite{poly}; even the dynamics of growth models evolves according to similar differential equations in one dimension \cite{spohn_growth}. \\ 
In the case of harmonic interactions the system is integrable, and it can be explicitly decoupled into independent normal modes. If a small nonlinearity is included in the interactions, the system is supposed to become chaotic. In particular, if we excite only a small fraction of modes in the initial state, we expect transport of energy to the other proper frequencies. The analysis of time scales associated to this relaxation process is an intriguing problem, which has been the topic of a scientific debate for more than $60$ years (detailed reviews can be found in \cite{gavallotti, Bambusi2015}).\\
The results of the very first study on the system by E.~Fermi, J.~Pasta, S.~Ulam and M.~Tsingou were published in $1955$ \cite{fermi_pasta_ulam1955, dauxois}. The authors were interested in the time needed by the system to reach equipartition from out-of-equilibrium initial conditions, in case the nonlinear dynamics was close to the integrable one. They therefore chose a two body-potential consisting of a main quadratic term, plus small higher-order corrections; they then assigned the total energy to the normal mode with the lowest frequency, and integrated numerically the equations of motion. To their surprise they found that, below a certain threshold, energy was exchanged only within a small subset of long wavelength modes. This unexpected result turned out to be in apparent contrast with the classical results from Ergodic Theory: Fermi himself in \cite{fermipoincare1} tried to generalize a Poincar\'e theorem, which was showing how an arbitrarily small perturbation of an integrable Hamiltonian suffices to ensure equipartition of energy. An intense and fruitful debate followed, leading to the interpretation of the observations within the framework of the KAM Theorem for discrete lattices \cite{Rink2006}. In the continuum limit the quasi-integrability behavior was shown to be related to the energy conservation of solitary waves \cite{Friesecke2014}. In general, the relation between conservation laws and localization phenomena in the system turns out to be a quite involved issue \cite{hajnal}.  Sixty year after its original formulation, the ``FPU Problem'' is still an open and relevant challenge for the foundation of the Dynamical Systems and Statistical Mechanics. \\
In the typical studies on the FPU Model, initial data are chosen very far from thermal equilibrium. However, it has been shown in \cite{caratieq} that the dynamics of the system can relax to metastable states even by sampling initial conditions with a large probability w.r.t. the Gibbs measure. The relevant observables in this case are the energy fluctuations, which can relax to nonequilibrium values. Other critical observables for the analysis of the system in canonical equilibrium are the energies of packets of normal modes. In \cite{maiocchi2013} is has been shown how to construct an adiabatic invariant from those variables ; in \cite{maiocchi} it is proven that localization translates into a non-exponential decay of the correlation function of those energies. The general procedure described there is based on the analysis of the poles of the Laplace transform of the correlation function. This can be estimated via the calculation of high order Lie derivatives of dynamical variables. Such quantities will also play a central role in the present work. \\
Another relevant issue involves the strength of the non-linearity: as the original localization has been observed in a quasi-harmonic potential, it is not obvious that Ergodicity violation can also be seen in the strongly anharmonic regime. This problem has been tackled in \cite{carati2015}. In this work the dynamics of the FPU chain is analyzed with highly nonlinear interactions. In this case the model can be interpreted as a glass former, as many disordered metastable basins appear in the potential landscape. For sufficiently low energy the dynamics remains trapped in one of those minima, and traces of this localization can be seen in an the anomalous relaxation of the harmonic part of the total energy.  \\
From the discussion above, it emerges that the analysis of phase space localization and quasi-integrability in the model is strongly dependent on the choice of a few relevant observables. We can in general refer to such dimensionality reduction procedures as coarse-graining. The most general and successful formalism in this context has been developed by Mori \cite{mori} and Zwanzig \cite{zwanzig}, and extended by Grabert \cite{grabert}. Within this framework one can write an exact equation of motion for the time evolution of one or few observables and the related self-time correlations. This kind of coarse-grained dynamics is governed by integro-differential equations in time, meaning that the current state of the system is history-dependent. This feature is a consequence of the fact that the Markovianity of Hamilton's (microscopic) equations is lost due to the severe dimensionality reduction from the entire phase space of the problem to the few variables considered. The Mori-Zwanzig formalism has been widely and successfully applied in Liquid-State Theory for the analysis of scattering functions, i.e.~correlations of the Fourier transforms of the density fluctuations. The strong interest behind this class of function is due to both their experimental availability and their effectiveness in capturing arrested states and glass transitions in amorphous liquids. For instance, a recursive approach for the reconstruction of the dynamic structure factor of the harmonic chain is discussed in \cite{wierling}. There, an iterative method is derived in order to reconstruct systematically the infinite fraction associated to the Langevin equation of the density fluctuations in Laplace space. The analysis is based on the extraction of static recurrants related to the moments of the observable.\\ \\
In the present work we analyze the dynamics of a generalization of the self-ISF in the FPU Model at equilibrium. The model and the observable are described in Section \ref{model}. In Section \ref{numerics} we present and discuss the results from Molecular Dynamics (MD) simulations. In particular we show how the relaxation of the ISF is affected by the temperature of the system, and how the integrable limit of the harmonic chain can be traced at very low temperatures. In Section \ref{constr} we describe a recursive algorithm for the exact calculation of the Taylor coefficients of the ISF. Via this method we can in particular avoid the numerical errors introduced by the calculation of time derivatives from MD simulations. At the end of that section we present a comparison between different approaches for the calculations of these moments, and a discussion of their temperature dependence. Finally, in Section \ref{gle} we show how we can reconstruct for a certain temperature range the whole memory kernel of the GLE of the scattering function. The method relies on a coarse-graining procedure derived via projection operators techniques in \cite{hugues}. Via the analysis of the first series coefficients we are moreover able to quantify the temperature regime associated to the observed dynamical crossover. The most technical aspects of the work have been collected in a series of appendices for the interested readers.

  \section{The Model}\label{model}
Consider a one-dimensional chain of $N$ identical particles with mass $m$ each, interacting through a nearest-neighbor non-linear potential. The Hamiltonian of the system can be written as:
\begin{equation*}
H = \sum_{j=0}^{N-1} \f{p_j^2}{2m} + \sum_{j=1}^{N-1} V_{\eta}(q_j-q_{j-1})
\end{equation*}
We set boundary conditions of Dirichlet type  
\begin{equation*}
q_0 = 0 \quad , \quad
q_{N-1} = L
\end{equation*}
where the total length $L$ is the sum of distances between neighboring particles. We use a quartic nonlinear potential with
\begin{equation}\label{pot}
V_{\eta}(r) = \alpha\Le[\Le(\frac r\sigma\Ri)^2+A(\eta)\Le(\f r\sigma\Ri)^3+B(\eta)\Le(\f r\sigma\Ri)^4\Ri]
\end{equation}
and coefficients
\begin{align*}
A(\eta) &= -2-4\eta \\
B(\eta) &= 1+3\eta \\
\end{align*}
This particular parametrization of the potential has been chosen such that the positions of the two minima remain fixed while $\eta=-V(\sigma)/\alpha$ tunes the unbalance between their depths (see Appendix \ref{app_pot}). 
\begin{figure}[H]
\begin{center}
\includegraphics[scale=0.6]{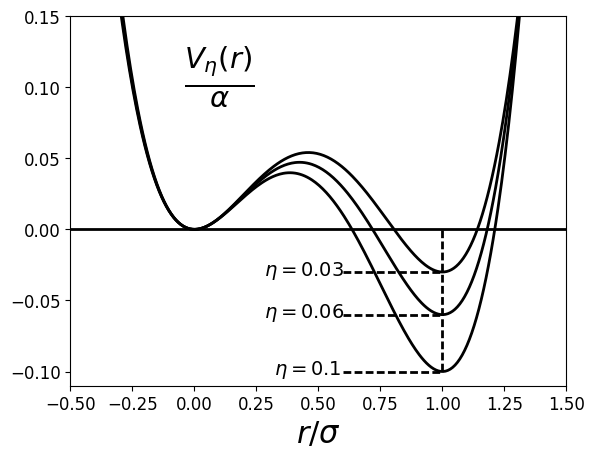}
\caption{Shape of the two body interaction potential for three different values of the parameter $\eta$}
\label{fig_pot}
\end{center}
\end{figure}
The main observable discussed in this work is the Intermediate Scattering function (ISF), constructed as follows: given the inter-particle distance between two tagged nearest neighbors $r_j\equiv q_j-q_{j-1}$, we can define its spacial density as 
\begin{equation*}
\rho^j(r) = \delta(r_j-r)
\end{equation*}
which can be Fourier transformed into the \textit{density fluctuation} 
\begin{equation*}
A_K^j = e^{iKr_j} 
\end{equation*} 
The time propagation of this variable is expanded as
\begin{equation}\label{time_df}
e^{iKr_j} \rightarrow e^{iKr_j(t)} =  e^{i\mc L t} e^{iKr_j} = \sum_n \f{(i\mc L t)^n}{n!} e^{iK r_j}
\end{equation}
where the Liouvillian $\mc L$ is defined as the operator acting on the phase space points $\bs \Gamma=(\mb q, \mb p)$ as
\begin{equation*}
i\mc L = \bs {\dot \Gamma}\cdot\f{\p}{\p\mb \Gamma} = \sum_{i=0}^{2N-2}\dot \Gamma_i\f{\p}{\p \Gamma_i} = \sum_{i=0}^{N-1} \Le[\dot q_i\f{\p}{\p q_i}+\dot p_i\f{\p}{\p p_i}\Ri] 	
\end{equation*}
and $e^{i\mc L t}$ is the time evolution operator under Hamilton's dynamics.
The magnitude of $1/K$ fixes the length-scale at which we probe the system. In the following we will stick to $\mc O(1/K)\sim\mc O (\sigma)$.
The time autocorrelation function of $A^j_K$ is defined as
\begin{equation}\label{defC}
C_j(t) = \int \mr d \bs \mu \;  \Le[ e^{i\mc L t} A^j_K(\bs \Gamma)\Ri] {A^j_K}^*(\bs\Gamma) = \lla e^{iK(r_j(t)-r_j(0))}\rra_\beta 
\end{equation}
where $\mr d \bs \mu = \mr d\bs \Gamma \; \rho_\beta(\bs\Gamma)$ is the Gibbs measure.  We omit in the notation the dependence of $C_j$ on the inverse temperature $\beta$ and the wavevector $K$, not to make the notation too heavy. The series expansion of $C_j(t)$ can be naturally deduced from the Taylor series of the time evolution operator:
\begin{equation}
C_j(t) = \sum_{n=0}^{+\infty} \f 1{n!} \lla \Le[(i\mc L t)^n e^{iKr_j}\Ri] e^{-iKr_j} \rra_\beta =  \sum_{n=0}^{+\infty} \f {(-1)^n}{(2n)!} \lla \lv(i\mc L t)^n e^{iKr_j}\rv^2 \rra_\beta = \sum_{n=0}^{+\infty} \f{t^{2n}}{(2n)!}\omega_{2n} \label{ck_tay}
\end{equation}
The second identity in eqn.~\ref{ck_tay} is a consequence of the anti-Hermicity of $i\mc L$ with respect to the scalar product induced by the canonical phase average. In the last identity we defined the coefficients
\begin{align}
\omega_{2n} &\equiv \lla \Le[ (i \mc L)^{2n} e^{i K r_j}\Ri] e^{-i K r_j} \rra_\beta =  (-1)^n \lla \lv(i\mc L )^n e^{iKr_j}\rv^2 \rra_\beta  \label{omega_def} \\
\omega_{2n+1} &\equiv 0 \;\;\; \forall n \ge 1 \label{null_odd}
\end{align}
The knowledge of the series of these coefficients would in principle allow to reconstruct the whole dynamics of $C_j(t)$ within the convergence radius of the series in eqn.~\ref{ck_tay} (a preliminary study on the convergence domain is discussed in Appendix \ref{cony_radius}). The solution of the correlation function for an ideal gas (with a purely kinetic Hamiltonian) can be computed exactly via the knowledge of the series of the respective $\Le\{\omega^{\free}_{2n}\Ri\}_{n=0}^{+\infty}$ (see \ref{app_free_dyn_corr}, eqn.~\ref{omega_id}):
\begin{equation}\label{c_free}
C^{\free}(t) = \sum_{n=0}^{+\infty}\f{t^{2n}}{(2n)!}\omega^{\free}_{2n} = \f 1 {\sqrt{\pi}} \sum_{n=0}^{+\infty} (-1)^n  \Le(\f {4K^2}{m\beta}\Ri)^n \f{(2n)!\sqrt{\pi}}{4^n n!}\f{t^{2n}}{(2n)!} = e^{-\f{ K^2 } {m\beta} t^2} = e^{-\Le(\f t\tau\Ri)^2}
\end{equation}
where we defined the timescale 
\begin{equation*}
\tau=\sqrt{\f{m\beta}{ K^2}}
\end{equation*}
that will ease the comparison of correlation functions with different values of $\beta$ and $K$ in the next section. \\
In case a potential term is introduced into the Hamiltonian, the dynamics is defined as \textit{mixing} if the time-correlation of any couple of dynamical variables converges to its phase average:
\begin{equation}\label{mix}
\lla f(t) g^*\rra_\beta - \lla f \rra_\beta \lla g^*\rra_\beta \to 0 \hspace{4mm}\text{for }t\to+\infty \hspace{10mm} \forall f, g\in L^2(\mr d \bs \mu)
\end{equation}
A mixing dynamical system is in particular \textit{ergodic}. In the case of our observable, eqn.~\ref{mix} reads as:
\begin{align}\label{relax_teo}
\lim_{t\to+\infty}C{\color{blue}_j}(t) &=\lim_{t\to+\infty} \lla e^{iKr_j(t)} {\Le(e^{iKr_j}\Ri)}^* \rra_\beta= \lla e^{iKr_j} \rra_\beta  \lla e^{iKr_j} \rra_\beta^* {\color{blue}}= \lla \cos(K r_j)\rra_\beta^2  +\lla \sin(K r_j)\rra_\beta^2 \nn \\
&\simeq \Le[\f 1{Z^1_r(\beta)}\int_{-\infty}^{+\infty}\mr d r_j e^{-\beta V_\eta(r_j)}\cos(Kr_j)\Ri]^2+ \Le[\f 1{Z^1_r(\beta)}\int_{-\infty}^{+\infty}\mr d r_j e^{-\beta V_\eta(r_j)}\sin(Kr_j)\Ri]^2\equiv C_{\beta} 
\end{align}
where 
\begin{equation*}
Z_r^1(\beta) = \int_{-\infty}^{+\infty} \mr d r  e^{-\beta V_\eta(r_j)}
\end{equation*}
The approximation in the eqn.~\ref{relax_teo} is due to the fact that statistical independence of positions and momenta is not strictly satisfied when fixing both the ends of the chain. To simplify the estimation of the phase averages in this case we will assume that such un-entanglement applies, as it would be the case of a chain with one free end. The approximation is justified for the short-times dynamics of a tagged degree of freedom far from the boundaries, such that the finite size effects are minimized. Let us remark that within the mentioned assumption the long-time limit $C_\beta$ becomes independent on $j$, according to eqn.~\ref{relax_teo}.\\
For the analysis of the decay to equilibrium of a correlation function, it is convenient to subtract the long time limit from $C_j(t)$ and to define the ISF  
\begin{equation*}
F_j(t) \equiv \f{C_j(t)-C_\beta}{C_j(0)-C_\beta}	
\end{equation*}
We again imply the dependence of $F_j$ on $K$ and $\beta$. We note that in the kinematic regime $F^{\free}(t)= C^{\free}(t)$ and that all the correlation functions considered here are bounded: as $A^j_K \in L^2(\mr d \bs \mu)$ we have
\begin{equation}
\llv \lla A_j^K(t)\Le(A_j^K\Ri)^*\rra_\beta\rrv \le \Le[\lla A_j^K(t) \Le(A_j^K(t)\Ri)^* \rra_\beta\lla A_j^K \Le(A_j^K\Ri)^*\rra_\beta\Ri]^{1/2} =1
\end{equation}
as a consequence of Schwartz's inequality. \\ \\
In the next section we present a numerical study of the dynamics of the ISF. The results from the numerical simulations show a continuous temperature driven transition from ideal gas to the harmonic limit as we initialize the system in cooler states; we discuss how this phenomenology is directly related to a localization of the dynamics in the phase space.
\section{Numerical Simulations}\label{numerics}
\begin{figure}[H]
	\centering
	\begin{minipage}{.5\textwidth}
		\centering
		\includegraphics[width=.9\linewidth]{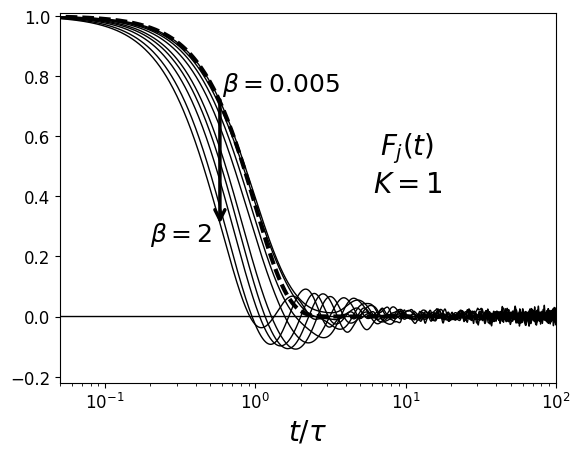}
		\caption{ISF for different values of $\beta$, $K=1$}
		\label{fig:ajk_diff_betas}
	\end{minipage}%
	\begin{minipage}{.5\textwidth}
		\centering
		\includegraphics[width=.9\linewidth]{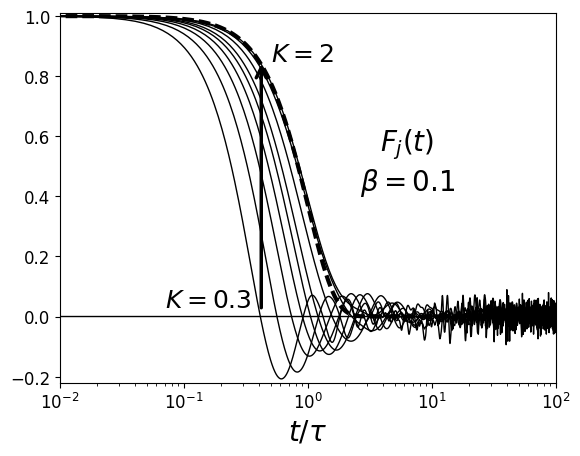}
		\caption{ISF for different values of $K$, $\beta=0.1$}
		\label{fig:ajk_diff_kappas}
	\end{minipage}
\end{figure}

\begin{figure}[H]
	\begin{center}
		\includegraphics[scale=0.6]{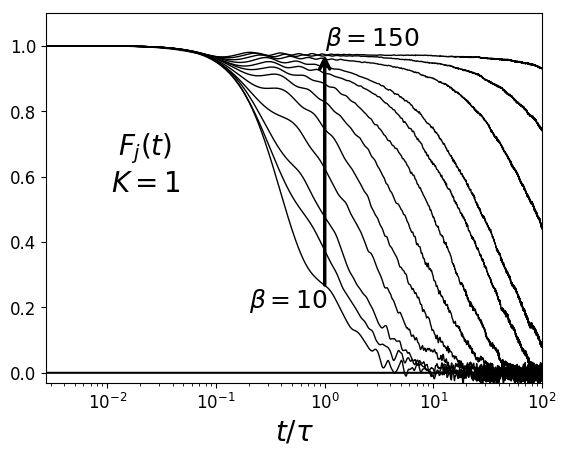}
		\caption{Time series of the ISF for lower temperatures than in Fig. \ref{fig:ajk_diff_betas} and \ref{fig:ajk_diff_kappas}. The wave vector is fixed in all the curves to $K=1$. As above, curves are plotted as a function of the temperature-rescaled time $t/\tau$.}\label{fig:ajk_high_betas}
		\label{fig:ajk_diff_betas_hi}
	\end{center}
\end{figure}

\begin{figure}[H]
	\begin{center}
		\includegraphics[scale=0.6]{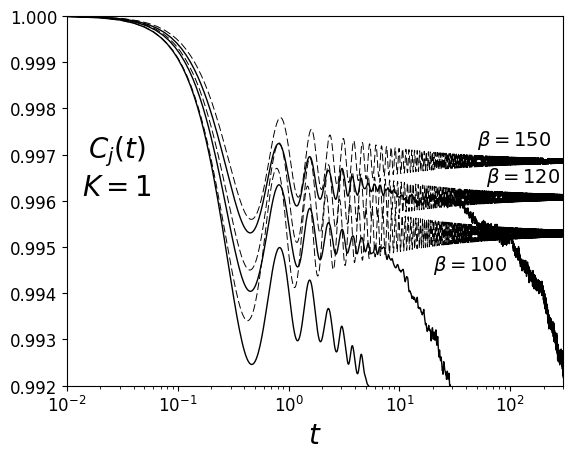}
		\caption{Time series of $C_j$ for the lowest temperatures. Dotted line: harmonic limits of the ISF (eqn.~\ref{Cht6}).}
		\label{ISF_lowT}
	\end{center}
\end{figure}

\begin{figure}[H]
	\begin{center}
		\includegraphics[scale=0.6]{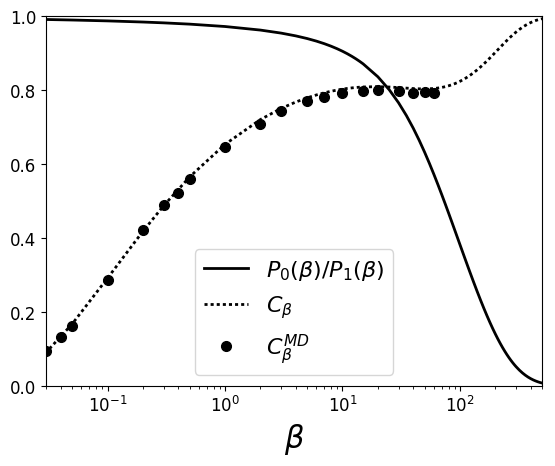}
		\caption{Solid line: probability ratio $P_0(\beta)/P_1(\beta)$; dotted line: phase average of the ISF; markers: long-time average of the ISF from the numerical simulations.  We fix here the wavevector to $K=1$.}\label{fig:transit}
	\end{center}
\end{figure}

To study the time evolution of the ISF, we run molecular dynamics simulations for a chain of $N=32,768$ particles with equal masses $m=1$. The parameters of the potential are set to $\alpha=9.5$, $\sigma=1$, $\eta=0.01$. In all the following figures the inverse temperature is given in units of the potential scale $\alpha$. The set of initial configurations is generated according to the Boltzmann distribution
\begin{equation*}
\rho_\beta(\mb r) = \f 1 {{Z(\beta)}} e^{- \beta\sum_{j=1}^{N-1} V_\eta(r_j)} = \f 1 {Z_1^{N-1}(\beta)}\prod_{j=1}^{N-1}e^{-\beta V_\eta(r_j)}
\end{equation*}
which we factorized into single particle distributions, each associated to the same partition function 
\begin{equation*}
Z_1(\beta) \equiv \int_{-\infty}^{+\infty} \mr d r e^{-\beta V_\eta(r)} 
\end{equation*}
The initial configurations of the particles are then determined by 
\begin{equation*}
q_j = \sum_{k=0}^j r_k, \hspace{10mm} 1\le j\le N-1
\end{equation*}
while the initial velocities are drawn from the Maxwell distribution at inverse temperature $\beta$. The trajectories of the particles $q_j$, $1\le j\le N-2$ are computed via the symplectic integration scheme \textit{leap-frog} \cite{allentid}, while keeping the first and last particle of the system fixed to their initial positions $q_0=0, \; q_{N-1}=L$.\\
The total length of the chain is a constant of motion at the trajectory level. However, expectation values w.r.t. the canonical distribution are computed by averaging over bundles of initial configurations for which the lengths $L$ are not identical -- but due to the large size of the system they are very close to a unique function of the temperature $L=L(\beta)$ .
\\ \\
Figure \ref{fig:ajk_diff_betas} shows the ISF as a function of the temperature-rescaled time $t/\tau$; each curve corresponds to a different value of $\beta$. Here and in the following graphs we show the dynamics of the central coordinate $j=\floor*{\f N2}$, the farthest from the boundaries of the system. The dotted line refers to the Gaussian relaxation in the limit of ideal gas, eqn.~\ref{c_free}. We can see that for increasing values of $\beta$ all the correlations converge faster (in the rescaled time) to their equilibrium value, while preserving a Gaussian decay. Figure \ref{fig:ajk_diff_kappas} shows the ISF for different wave vectors as a function of $t/\tau$. As $K$ is the inverse of the wavelength at which the system is probed, the system tends to equilibrate faster at smaller $K$, i.e. the details of the microscopic dynamics are not probed on larger length-scales.
Fig. \ref{fig:ajk_high_betas} shows the time series of the ISF for values $\beta$ higher than the ones considered in Fig. \ref{fig:ajk_diff_betas}. In this regime the equilibration time increases continuously for decreasing temperatures, in contrast to Fig. \ref{fig:ajk_diff_betas}. We can interpret this process as a dynamical localization transition exhibited in the temperature range where the details of the shape of the two-body potential become relevant. The particles' dynamics deviates from the ideal gas limit as the system becomes more and more confined to the lower potential wells. Due to energy constraints the particle's kinetic energy is in general not sufficient for them to jump over the barrier in the two body interaction in Fig. \ref{fig_pot}. Only large fluctuations may allow this hopping process to happen, thus they occur at a vanishing rate for $\beta\to+\infty$.
Now we focus on the time dependence of the three highest curves in Fig.~\ref{fig:ajk_diff_betas_hi}, corresponding to the lowest temperatures considered. In Fig.~\ref{ISF_lowT} we compare these correlation functions with the harmonic limits for the same values of $\beta$ and $K$ (dotted lines); a semi-analytical expression for this limit is derived in Appendix \ref{harm_lim} by linearizing the potential around its global minimum. We note that the initial decay and the early oscillations between the analytical and numerical curves remain close for finite times; the anharmonic contributions finally drive the correlation towards their equilibrium value, in agreement with eqn.~\ref{relax_teo}.
For increasing values of $\beta$, the distribution of the inter-particle distances becomes more and more localized in the global minimum of the potential in $r=\sigma=1$ (see Fig. \ref{fig_pot}). We can quantify this temperature-driven localization by computing the probabilities
\begin{align*}  
P_0(\beta) &= \int_{-\infty}^{r_c} \mr d r \; \rho_\beta(r) \\
P_1(\beta) &= \int_{r_c}^{+\infty} \mr d r \; \rho_\beta(r)
\end{align*}
where $r_c$ is the central local maximum in the two body potential, and of course $P_0(\beta)+P_1(\beta)=1$. We can see in Fig. \ref{fig:transit} that the probability ratio $P_0/P_1$ is a monotonically decreasing function of $\beta$. In the same picture are also displayed the long time limit of the correlation function $C_\beta$ computed with the theoretical phase average in eqn.~\ref{relax_teo} (dotted line) and by extracting the limits $C^{MD}_\beta$ from the numerical simulations (dots). We can see that the main drop of the probability ratio happens to be the range $ 10 \leq \beta \leq 10^2$, in accordance with the non-monotonic behavior of $C_\beta$. This temperature range corresponds to the qualitative change of behavior observed in the curves in Fig. \ref{fig:ajk_diff_betas} and \ref{fig:ajk_high_betas}. 

In this section we showed by numerical evidence that the ISF in the present generalization is a valuable tool for the analysis of the dynamics and related localization phenomena in the FPU chain.     
The evolution of the function is sensitive to the temperature at which the system is initialized, and a critical value $\beta=\beta_c \sim 10/\alpha$ could be identified as a separator between two behaviors: the fast quasi-exponential decay in Fig. \ref{fig:ajk_diff_betas}, and a strongly non-Gaussian relaxation in Fig. \ref{fig:ajk_high_betas} converging towards the harmonic limit in Fig. \ref{ISF_lowT}. \\
In the following we discuss a method for the recursive calculation of the series coefficients of the ISF. We will then use these coefficients to study the memory effects exhibited by the system.

\section{Construction of the Dynamics}\label{constr}
In this section we present an iterative method for the exact reconstruction of the ISF via the calculation of the correlators $\omega_n$, as defined in eqn.~\ref{omega_def}. This approach is based on a direct evaluation of the action of $(i\mc L)^n$ on the density fluctuations. \\ When dealing with the full microscopic dynamics, we need to choose a proper set of coordinates to keep track of the coupling between the different d.o.f., and to reduce the complexity of the many-body problem. As the potential depends only on the distances between nearest-neighbors (eqn.~\ref{pot}), the displacements $r_j=q_j-q_{j-1}$ appear to be a natural set of configurations. One can show that $\pi_j=\sum_{n=j}^{N-1}p_j$ are the conjugate momenta to these configurations. However, due to the many body sum in this expression, the calculation of phase averages of functions of $\pi_j$'s becomes rather involved; it is much simpler for our analytical and numerical purposes to use the non-canonical set of coordinates $(\mb r, \mb p)$.  As shown in Appendix \ref{change_liou} that action of the Liouville operator in these coordinates is rewritten as
\begin{align}\label{liou}
i\mc L & =\sum_{i=1}^{N-1} \Le[-\f{\p V(r_i)}{\p r_i}\Le(\f{\p }{\p p_i}-\f{\p}{\p p_{i-1}}\Ri) + \f 1{m}\Le(p_i - p_{i-1}\Ri)\f{\p}{\p r_i} \Ri]
\end{align}	
The action of the $(i\mc L)^n$ on the density fluctuations can in general be written as the following complex polynomial in the coordinates
\begin{equation}\label{general_exp}
(i\mc L)^n e^{iKr_j} \equiv \sum_{\substack{m_{k_{\min}},\cdots, m_{k_\max} \\ s_{l_{\min}},\cdots, s_{l_\max} }}
\mc I^{(n)}_{\substack{m_{k_{\min}},\cdots, m_{k_\max} \\ s_{l_{\min}},\cdots, s_{l_\max} }} \Le(r_{k_\min}^{m_{k_\min}}\cdots r_{k_\max}^{m_{k_\max}} \Ri)\Le(p_{l_\min}^{s_{l_\min}}\cdots p_{l_\max}^{s_{l_\max}} \Ri)e^{iKr_j}\equiv \sum_{\mb{ms}} \mc I^{(n)}_{\mb m, \mb s} \mb r^{\mb m}\mb p^{\mb s} e^{iKr_j}
\end{equation}
with complex coefficients $\mc I^{(n)}_{\mb{ms}} = \mc  I^{(n, \Re)}_{\mb{ms}} + i\mc I^{(n, \Im)}_{\mb{ms}}$. It would be computationally difficult to tackle eqn.~\ref{general_exp}  in case all the degrees of freedom of the system entered in the expansion at any order, i.e. if the sum ran over $\mc O(2N)$ indexes. Conveniently, the interaction at the lowest orders in $n$ is restricted to a relatively small subset of coordinates around the tagged $j$-th degree of freedom. This is a direct consequence of the nearest-neighbor nature of the interaction. In the following we will formalize this observation. From the expansion of $(i\mc L)^n e^{iKr_j}$ we can identify relevant products of differentials that control the propagation of the interaction from $r_j$ to any other d.o.f. of the chain; we will refer to these contributions as \textit{spreading operators}. Via these cross derivatives we can perform a systematic study of the extent of the non-locality as a function of the order $n$. For example, the composition of derivatives connecting the configuration $r_j$ of the tagged particle to another generic coordinate $r_k$ located on the right (R) of $r_j$ ($k>j$) is given by
\begin{align}
&{\sigma_r^R}(n, j, k-j ) \equiv \theta\Le(\floor*{\f n2}-k+j\Ri) \Le[ \Le( -\f{\p V}{\p r_k} \Ri)\Le(\f{\p}{\p p_k}-\f{\p}{\p p_{k-1}}\Ri)\Ri]^{\theta(n-2)}\times \nn \\
&\times\Le\{ \prod_{i=0}^{ k-j - 2} \Le[ \f 1{m}(p_{j+ i+1}-p_{j+i}) \f{\p}{\p r_{j + i + 1}}\Ri] \Le[ \Le( -\f{\p V}{\p r_{j+i+1}} \Ri)\Le(\f{\p}{\p p_{j+ i+1}}-\f{\p}{\p p_{j+i}}\Ri)\Ri]\Ri\}^{\theta\Le(n-4\Ri)}\Le[\f 1{m}(p_j-p_{j-1})\f{\p}{\p r_j}\Ri]^{\theta \Le(n-2 \Ri)} \label{sigmarR} 
\end{align}
where 
\begin{equation*}
\theta(x) =\begin{cases}
1,& x\ge 0 \\
0, & x<0
\end{cases}
\end{equation*}
Fox fixed $n$ and $j$, the prefactor $ \theta\Le(\floor*{\f n2}-k+j\Ri)$ reduces the {effective domain of $\sigma_r^R$ to the subset of coordinates that enter in $(i\mc L)^n e^{iKr_j}$ i.e., that participate in the propagation of the density fluctuations at order $n$. In Appendix \ref{app_spread} we show the symmetric structure of $\sigma^L_r$, connecting $r_j$ to another coordinate $r_k$ on the left $(L)$ of $j$ ($k<j$), as well as the expressions of the analogous operators for the momenta ${\sigma_p^{(L/R)}}$. Via the spreading operators it becomes straightforward to list the subset of neighboring distances entering the interaction at order $n$:
\begin{align}\label{Kr}
K_r(n, j) &\equiv \Le\{j-\floor*{\f{n}2},\cdots, j+\floor*{\f{n}2} \Ri\} \in \mathbb N^{\Le(2\floor*{\f n 2}+1\Ri)} \equiv \Le\{k_{\min},\cdots, k_\max\Ri\}
\end{align}
An analogous set for the momenta are given in eqn.~\ref{Lp}. The boundaries of the sums in eqn.~\ref{general_exp} can again be determined identifying of the relevant differential operators of $(i\mc L)^n$. We can define $\mc M_r(n, j,\lv j-k\rv)$ and $\mc S_p(n, j, \lv j-l\rv)$ as the leading power respectively for a configuration $r_k$ and a momentum $p_l$ in the sum. In vector notation these definitions read as:
\begin{equation*}
\mb m = \begin{pmatrix}  m_{k_{\min}} \\\vdots \\ m_{k_{\max}} \\  \end{pmatrix} \le  \begin{pmatrix} \mc M_r(n, j, j-k_{\min} ) \\\vdots \\ \mc M_r(n, j, k_{\max}-j) \end{pmatrix}\equiv \mb M
\end{equation*}
and analogously for the momenta
\begin{equation*}
\mb s = \begin{pmatrix} s_{l_{\min}} \\\vdots \\ s_{l_{\max}} \end{pmatrix} \le  \begin{pmatrix} \mc S_p(n, j, j-l_{\min} ) \\\vdots \\ \mc S_p(n, j, l_{\max}-j) \end{pmatrix}\equiv \mb S
\end{equation*}
The boundaries of the sums in equation eqn.~\ref{general_exp} are therefore fixed to
\begin{align}
(i\mc L)^n e^{i K r_j}= \sum_{\substack{\mb 0 \le\mb m\le \mb M \\\mb 0 \le\mb s\le \mb S}}\mc I^{(n)}_{\mb{ms}}\mb{r^m p^s} e^{i K r_j} \equiv \Le[p_n^\Re(\mb r, \mb p)+i p_n^\Im(\mb r, \mb p)\Ri]e^{iKr_j}
\end{align}
where we additionally split the real and imaginary part of the polynomial expansion in the terms $p_n^\Re(\mb r, \mb p)$ and $p_n^\Im(\mb r, \mb p)$. By computing the averages over the Gibbs distribution we obtain as a the final expression of the Taylor coefficients
\begin{align}\label{exp_fin}
\omega_{2n} &= (-1)^n \lla \lv (i\mc L)^ne^{iKr_j} \rv^2 \rra_\beta = (-1)^n \lla \lv p_n^\Re(\mb r, \mb p)+i p_n^\Im(\mb r, \mb p) \rv^2 \rra_\beta = (-1)^n \Le[ \lla (p_n^\Re(\mb r, \mb p))^2\rra_\beta + \lla (p_n^\Im(\mb r, \mb p))^2\rra_\beta \Ri] = \nn \\
&=(-1)^n \sum_{\substack{\mb 0 \le\mb{m,m'}\le\mb M \\\mb 0 \le\mb{s,s'}\le \mb S}} \Le( \mc I^{(n,\Re)}_{\mb{ms}} \mc I^{(n,\Re)}_{\mb{m's'}}+ \mc I^{(n,\Im)}_{\mb{ms}} \mc I^{(n,\Im)}_{\mb{m's'}} \Ri) \lla \mb{r^{m+m'}p^{s+s'}}\rra_\beta 
\end{align}
Note that, as a consequence of the assumption of stationarity, knowledge of the $n$-th order tensor of coefficients $\mc I^{(n)}$ suffices to calculate $\omega_{2n}$. We present now a recursive method to compute the entries of $\mc I^{(n)}$. We can extract a relation $\mc I^{(n)}\rightarrow \mc I^{(n+1)}$ directly from the action of the powers of $i\mc L$ on the dynamical variable: 
\begin{align*}
(i\mc L)^{n+1}e^{iKr_j} = \sum_{\substack{\mb 0 \le\mb m\le \mb M \\\mb 0 \le\mb s\le \mb S}}\mc I^{(n+1)}_{\mb{ms}}\mb{r^m p^s} e^{i K r_j} = (i\mc L)(i\mc L)^n e^{iKr_j} = i\mc L \Le(\sum_{\substack{\mb 0 \le \mb m\le \mb M \\\mb 0 \le \mb s\le \mb S}}\mc I^{(n)}_{\mb{ms}}\mb{r^m p^s} e^{i K r_j}\Ri)
\end{align*}
As an example, we can isolate from $i\mc L$ the contribution of 
\begin{equation*}
i\mc L_p^\gamma\equiv-\f{\p V(r_\gamma)}{\p r_\gamma}\Le(\f{\p }{\p p_\gamma}-\f{\p }{\p p_{\gamma-1}}\Ri) = -\f{\alpha}{\sigma^2}\Le(2 r_\gamma+3A(\eta)\f{r_\gamma^2}\sigma + 4B(\eta)\f{r_\gamma^3}{\sigma^2}\Ri)\Le(\f{\p }{\p p_\gamma}-\f{\p }{\p p_{\gamma-1}}\Ri)
\end{equation*}
with $k_\min \le\gamma\le k_\max$ and $l_\min+1\le \gamma \le l_\max$ (see eqn.~\ref{Lp}). We can then determine explicitly its action on the monomials in the expansion at order $n$:
\begin{align*}
&i\mc L_p^\gamma\Le[ \Le( r_{k_{\min}}^{m_{k_{\min}}} \cdots r_\gamma^{m_\gamma} \cdots r_{k_{\max}}^{m_{k_{\max}}} \Ri)\Le( p_{l_{\min}}^{s_{l_{\min}}} \cdots p_{\gamma-1}^{s_{\gamma-1}} p_\gamma^{s_\gamma} \cdots p_{l_{\max}}^{s_{l_{\max}}} \Ri) \Ri]= \\
&=-\f{\alpha}{\sigma^2}\Le[2 r_\gamma+3A(\eta)\f{r_\gamma^2}\sigma + 4B(\eta)\f{r_\gamma^3}{\sigma^2}\Ri]\Le(\f{\p }{\p p_\gamma}-\f{\p }{\p p_{\gamma-1}}\Ri) \Le[ \Le( r_{k_{\min}}^{m_{k_{\min}}} \cdots r_\gamma^{m_\gamma} \cdots r_{k_{\max}}^{m_{k_{\max}}} \Ri)\Le( p_{l_{\min}}^{s_{l_{\min}}} \cdots p_{\gamma-1}^{s_{\gamma-1}} p_\gamma^{s_\gamma} \cdots p_{l_{\max}}^{s_{l_{\max}}} \Ri) \Ri]= \\
&=  -\f{\alpha}{\sigma^2}\Le[r_{k_{\min}}^{m_{k_{\min}}}\cdots\Le(2r_\gamma^{m_\gamma+1}+ 3A(\eta)\f {r_\gamma^{m_\gamma+2}}{\sigma} +  4B(\eta)\f{r_\gamma^{m_\gamma+3}}{\sigma^2}\Ri)\cdots r_{k_{\max}}^{m_{k_{\max}}}\Ri]\Le[p_{l_{\min}}^{s_{l_{\min}}} \cdots \Le(p_{\gamma-1}^{s_{\gamma-1}} s_\gamma p_\gamma^{s_\gamma-1}-s_{\gamma-1} p_{\gamma-1}^{s_{\gamma-1}-1} p_\gamma^{s_\gamma}\Ri) \cdots p_{l_{\max}}^{s_{l_{\max}}}\Ri]\\
\end{align*}
The contribution of $i\mc L_p^\gamma$ to $\mc I^{(n+1)}_{\mb{ms}}$ is then given by
\begin{equation}\label{ilpgamma}
{\mc I}^{(n+1)}_{\mb{m,s}}\Big\vert_{p,\gamma} \equiv \alpha\sum_{k=0,1}(-1)^{k+1} s_{\gamma-k} \Le[\f 2{\sigma^2}\mc I^{(n)}_{\mb{m}+\mb{\hat e}_\gamma,\mb s-\mb{\hat e}_{\gamma-k}}+\f{3}{\sigma^3}A(\eta)\mc I^{(n)}_{\mb{m}+2\mb{\hat e}_\gamma,\mb s-\mb{\hat e}_{\gamma-k}}+\f{4}{\sigma^4}B(\eta)\mc I^{(n)}_{\mb{m}+3\mb{\hat e}_\gamma,\mb s-\mb{\hat e}_{\gamma-k}}\Ri]
\end{equation}
where the unit vectors $\mb e_j$ are defined via their action on vectors in any dimension $\mathbb R^n$ such that: 
\begin{equation*}
\mb v + \alpha \mb e_j = \Le\{v_1, \cdots, v_j +\alpha, \cdots, v_n\Ri\},\hspace{5mm}\forall c\in \mathbb R
\end{equation*}
The other terms of the recursion relation are derived in Appendix \ref{app_recursion}, while in Appendix \ref{map_tensor} we present a general method for the numerical management of high dimensional tensors as the ones considered in the procedure. \\ \\
In Fig.~\ref{fig:dynCorr} we show the dynamical correlators computed with different methods, normalized by the analytically known values of the ideal gas (eqn.~\ref{omega_id}). The numerical coefficients are determined by the calculation of high order time derivatives in the trajectories of the density fluctuation, according to 
\begin{equation*}
\omega_{2n}^{MD} \equiv\lla \Le.\f{\mr d^{2n}}{\mr d t^{2n}}  e^{iKr_j(t)}\Ri|_{t=0}e^{-iKr_j} \rra_\beta 
\end{equation*}
We observe optimal agreement between the recursive approach and the numerical values. The magnitude of $\omega_2$ is the same in the kinematic and the dynamic regime (and is therefore analytically known); the analytical value of $\omega_4$ in presence of nonlinearities is computed in the Appendix \ref{app_sn_dyn_corr}.
\begin{figure}[H]
 	\begin{center}
 		\includegraphics[scale=0.6]{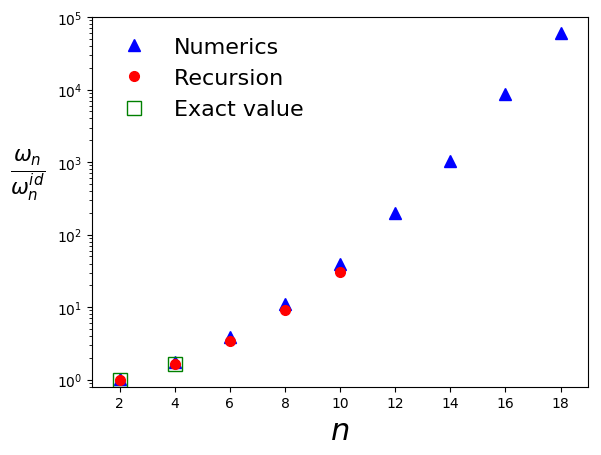}
 		\caption{$\omega_n/\omega_n^{\free}$ from the MD simulations (triangles) and via the recursion relation derived in the work (circles); comparison with the analytical result for $n=2, 4$ (see \ref{app_sn_dyn_corr}); we fix to $\beta=0.1$, $K=1$.}\label{fig:dynCorr}
 	\end{center}
\end{figure}
In Fig.~\ref{omega_diff_betas_id} we show the sequence of the series coefficients computed via the recursion relation, for a wide range of inverse temperatures. We notice for decreasing values of $\beta$ a convergence of the sequences to the ideal limit. In Fig.~\ref{omega_diff_betas_harm} we present the same series of coefficients $\omega_n$ normalized w.r.t. the correspondent values computed in the harmonic limit $\omega_n^h$. The convergence of the ratios to unity is not as neat as in the previous case. We ascribe the discrepancy to the fact that a few approximations enter the semi-analytical estimate of the harmonic limit of the ISF, as discussed in Appendix~\ref{harm_lim}; in particular the system is far from its continuum limit.
\begin{figure}[H]
	\centering
	\begin{minipage}{.5\textwidth}
		\centering
		\includegraphics[width=.9\linewidth]{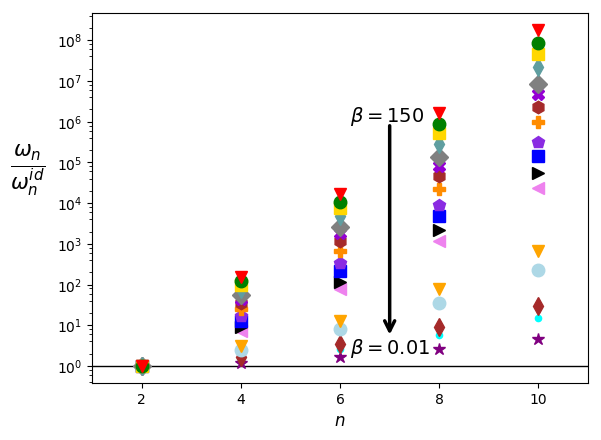}
		\caption{$\omega_n/\omega_n^{\free}$ for different values of $\beta$ and $K=1$}
\label{omega_diff_betas_id}
	\end{minipage}%
	\begin{minipage}{.5\textwidth}
		\centering
		\includegraphics[width=.9\linewidth]{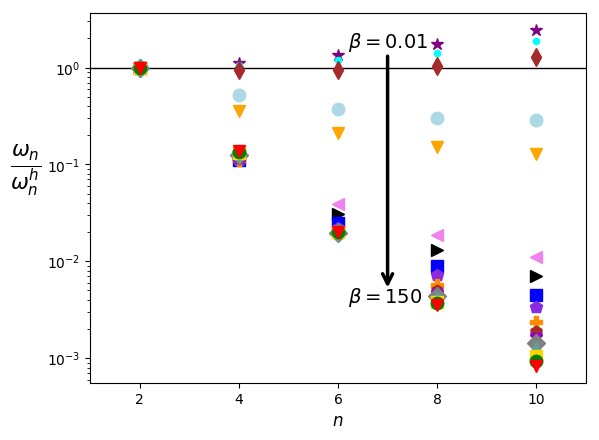}
		\caption{$\omega_n/\omega_n^h$ for different values of $\beta$ and $K=1$}
		\label{omega_diff_betas_harm}
	\end{minipage}
\end{figure}
Via the procedure presented in this section it is possible to compute the first orders of the dynamics of the ISF exactly; higher order Taylor coefficients could be computed with increasing complexity by computer algebra. Instead of pushing in this direction, in the next section we will describe a combined numerical and analytical approach that will enable us to study the dynamics of the ISF with no need of additional effort in the compuations. We will in particular determine the memory effects encapsulated in the evolution of this relevant correlation function. \\
\section{The Generalized Langevin Equation}\label{gle}	
The dynamics of the non normalized ISF at equilibrium is governed by the GLE
\begin{equation}\label{ca_stat}
\f{\mr d C_j(t)}{\mr d t} = \int_0^{t} \mr d t' K_j(t-t')C_j(t')
\end{equation}
which can be derived via projection operator techniques \cite{reichman}. 
}
The memory kernel $K_j$ controls to which extent the state of the process at a certain instant in time depends on the previous evolution. In the Markovian (memory-less) limit $K_j(t)=-\gamma \delta(t),\; \gamma\in \mathbb R^+$, i.e.~the solution of eqn.~\ref{ca_stat} decays exponentially. We can see from the results of Section $3$ that the dynamics of the ISF is far from exponential; it is therefore nontrivial to reconstruct the shape of $K_j$. In this section we recover the whole time series of the kernel for a certain range of temperatures and we define a criterion to quantify the ``strength of the memory''. \\

Following \cite{hugues} and \cite{hansenmd}, we can reconstruct systematically the memory of the process via a relation between the series coefficients of the kernel
\begin{equation*}
\kappa_n \equiv \Le(\left.\f{\mr d}{\mr d t}\Ri)^n K(t)\right|_{t=0}
\end{equation*}	
and the $\omega_n$ defined in eqn.~\ref{omega_def}. This correspondence for processes at equilibrium can be implicitly written as 
\begin{equation}\label{rec_kappa}
\kappa_{n} = \mc F_{n}(\omega_0, \omega_2, \cdots, \omega_{n+2})
\end{equation}
being $\mc F_{n}$ a nonlinear combination of its arguments. \\
The reconstruction of $K_j$ from its series expansion is reduced to the calculation of its even series coefficients, as the odd ones are identically null at equilibrium. This can be easily shown by inverting $t\rightarrow -t$ and $t'\rightarrow -t'$ in eqn.~\ref{ca_stat} and noticing that as $C_j$ is even it follows $K_j(t-t')=K_j(t'-t)$; this means that the series expansion of $K_j$ must only include even contributions.

\begin{figure}[H]
	\begin{center}
		\includegraphics[scale=0.6]{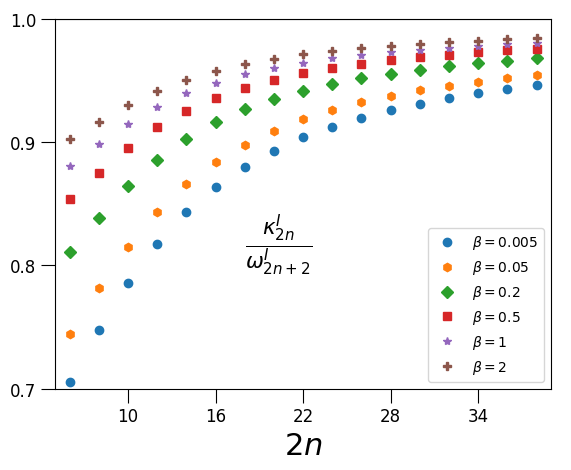}
		\caption{Ratio $\kappa^I_{2n}/\omega^I_{n+2}$ as a function of $n$, for different values of the inverse temperature $\beta$}\label{conv_kap}
	\end{center}
\end{figure}

\begin{figure}[H]
	\begin{center}
		\centering
\includegraphics[scale=0.6]{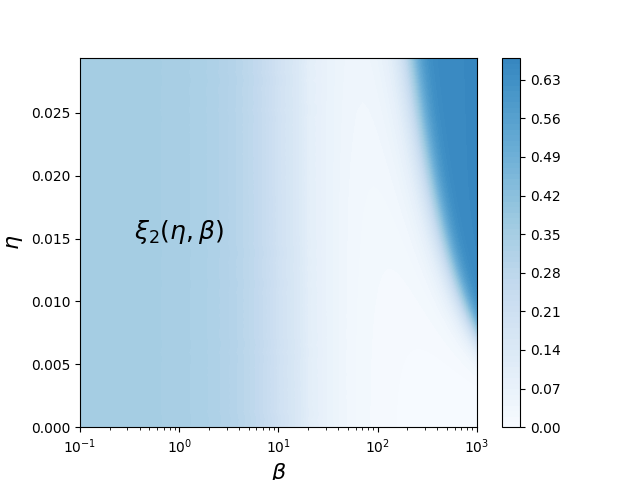}
\caption{$\xi_2$ for different values of $\beta$ and $\eta$, $K=1$}\label{x2_fig}
	\end{center}
\end{figure}

The reconstruction of the time series of the kernel via eqn.~\ref{rec_kappa}, requires the knowledge of the high-order series coefficients of $C{\color{blue}_j}(t)$. The calculation of these terms can be simplified by introducing an interpolating approximation for the correlation function $C^I_j\in C^{\infty}(\mathbb R^+)$, such that its derivatives in $t=0$ provide an estimate of the required Taylor coefficients. To determine a proper functional expression for the interpolating function we can proceed as follows. We can see from the main plot in Figure \ref{fig:ajk_diff_betas} that for sufficiently high temperatures (for $\beta \lesssim 2$) the initial decay of $F_j(t)$ preserves the Gaussian shape of the ideal gas (eqn.~\ref{c_free}). It appears therefore convenient in this low-$\beta$ regime to define as \textit{ansatz}:\\ 
\begin{align}\label{ans}
C_j^I(t) &\equiv (1-C_\beta)e^{-at^2}+C_\beta 
\end{align}
where the superscript $I$ stands for ``interpolation''. The long time limit $C_\beta$ is fixed by eqn.~\ref{relax_teo}, while the Gaussian parameter $a=a(\beta)$ can be extracted from the time series of $C_j(t)$ in the numerical simulations: 
\begin{equation}\label{param_a}
a= \f1{2(C_\beta-1)}\left. \f{\mr d^2 C_j(t)}{\mr d t}\right|_{t=0}
\end{equation} 
Via eqn.~\ref{param_a} we can quantify the deviation from time-temperature superposition in presence of the interaction. (At this level of description we explicitly choose to neglect the oscillatory patterns exhibited in the curves in Figures \ref{fig:ajk_diff_betas} and \ref{fig:ajk_diff_kappas} before the final relaxation; we can interpret these as traces at finite temperature of the harmonic dynamics of the ISF, according to eqn.~\ref{Cht6}). 
\\ \\ 
Let us remark that the shape of the \textit{ansatz} in eqn.~\ref{ans} is compliant with the requirement that the odd series coefficients of the ISF are identically null. We recall from eqn.~\ref{ck_tay} that this restriction is a direct consequence of the property of anti-Hermicity of the $i\mc L$ at equilibrium. An interpolation of the ISF with a general stretched exponential
\begin{equation}\label{stre_exp}
f(t) = (1-A)e^{Bt^\alpha}+A, \hspace{5mm}\alpha \neq 2
\end{equation}
would violate this symmetry. These functions are frequently used to fit slowly decaying correlations \cite{stretched_exp}. The applicability of the \textit{ansatz} in eqn.~\ref{stre_exp} is however restricted to regimes where the Liouville operator $\mc L$ is not Hermitian, as it happens in general out-of-equilibrium processes. \\ \\
With the \textit{ansatz} eqn.~\ref{ans} we can compute at any order the Taylor coefficients of $C^I_j$:
\begin{align}
\omega^I_{2n} &\equiv \left. \f{\mr d^{2n} C{\color{blue}^I}_j(t)}{\mr d t^{2n}}\right|_{t=0} = (1-C_\beta)a^{n} \left. \Le(\f{\mr d}{\mr d (\sqrt a t)}\Ri)^{2n} e^{-(\sqrt a t)^2}\right|_{t=0} + \delta_{n,0}C_\beta = \nn \\
&=\left. (1-C_\beta) a^n (-1)^{2n} H_{2n} (\sqrt a t) e^{-(\sqrt a t)^2} \right|_{t=0} + \delta_{n,0}C_\beta =(1-C_\beta) a^n H_{2n} + \delta_{n,0}C_\beta \label{om_i}
\end{align}
where we introduced $H_n(x)$, the $n$-th Hermite polynomials, and the Hermite numbers $H_n\equiv H_n(0)$. From the recursion relation $H_{n+1}(x) =2xH_n(x)-2nH_{n-1}(x)$ and the initial conditions $H_0(0)=1, H_1(0)=0$, we can write the closed form
\begin{align*}
H_n = 
\begin{cases}
(-2)^{n/2}(n-1)!!& \text{n even}\\ 
0& \text{n odd}
\end{cases}
\end{align*}
We can insert the estimates of eqn.~\ref{om_i} into eqn.~\ref{rec_kappa} to compute the Taylor coefficients $\kappa_n^I$ of the memory kernel $K_j^I$ for $n \lesssim 40$. Above this order, the calculation of the exact expression of $\mc F_n$ become computationally expensive, due to the increasing combinatory complexity. A simplification is possible according to the theoretical arguments in Appendix \ref{high_ord_kappas}, where we show how that for higher orders of $n$ we can approximate
\begin{equation}\label{high_kap}
\kappa^I_{2n}\simeq\omega^I_{2n+2} \hspace{10mm} n\gtrsim 40
\end{equation}
This estimate is also confirmed numerically from the convergence to unity of the ratio $\kappa^I_{2n}/\omega^I_{2n+2}$, as shown in Fig. \ref{conv_kap}. \\ 
With this additional approximation we can now compute the short-time expansion the kernel
\begin{equation}\label{Knmax}
K_j^{I,N_\max}(t-t') \equiv \sum_{n=0}^{N_\max} \f{\kappa^I_{2n}}{(2n)!}(t-t')^{2n}
\end{equation}
till order $N_\max=180$. The results for $K_j^{I,N_\max}$ at different values of $\beta$ correspond to the first decays and the (dotted) long-time divergent lines in Fig. \ref{kers}. An estimate of the convergent decay at long time is derived in Appendix \ref{tail_ker}.
\begin{figure}[H]
	\begin{center}
		\includegraphics[scale=0.6]{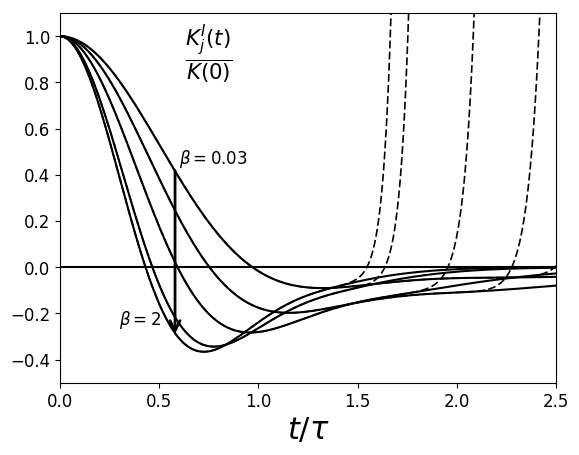}
		\caption{Reconstruction of the memory kernel of the GLE (eqn.~\ref{ca_stat}) for different values of $\beta$ and $K=1$, as a function of the rescaled time $\tau$. The plot of the bare expression in eqn.~\ref{Knmax} diverges at finite times (dashed lines); the extrapolation of the convergent limits (thick lines at long times) is discussed in Appendix \ref{tail_ker}.}\label{kers}
	\end{center}
\end{figure}
In the following we would like to define a criterion to quantify the memory effects at different temperatures, and to relate them to results from the MD simulation in Section \ref{numerics}. We can make a direct comparison between the timescales of the ISF and the one of the related kernel by expanding these two functions as second order:
\begin{align*}
F_j(t) &= \f 1{1-C_\beta}\Le[\sum_{n=0}^{+\infty} \f{\omega_{2n}}{(2n)!}t^{2n} -C_\beta\Ri] \equiv \f 1{1-C_\beta}\Le[\sum_{n=0}^{+\infty}(-1)^n\Le(\f{t}{\tau_{2n}^C}\Ri)^{2n}-C_\beta\Ri] = 1-\Le(\f t{\tau_{2}^C}\Ri)^2+\cdots \\
\f{K_j(t)}{K_j(0)} &=\f 1 {K_j(0)} \sum_{n=0}^{+\infty} \f{\kappa_{2n}}{(2n)!}t^{2n} \equiv  \sum_{n=0}^{+\infty}(-1)^n\Le(\f{t}{\tau_{2n}^K}\Ri)^{2n} = 1 - \Le(\f{t}{\tau_{2}^K}\Ri)^{2}+\cdots
\end{align*}
The parabolic approximation of $F_j$ and $K_j$ vanish respectively at the instants
\begin{equation*}
\tau_2^F = \sqrt{\llv\f{2(1-C_\beta)}{ \omega_2}\rrv} \hspace{8mm}\tau_2^K = \sqrt{\llv \f{2\omega_2}{\omega_4-\omega_2^2}\rrv}
\end{equation*}
We can therefore define the dimensionless ratio
\begin{equation*}
\xi_{2} = \xi_{2}(\beta, \eta) \equiv \Le(\f{\tau_{2}^K}{\tau_{2}^F} \Ri)^2= \f 1{1-C_\beta} \f{\omega_2^2}{ \omega_{4}-\omega_2^2}
\end{equation*}
quantifying the extent of the timescale of the memory kernel w.r.t. the one of the correlation. 
This function of $\beta$ and $\eta$ is plotted in Fig. \ref{x2_fig}. In the limit $\beta \to 0$ the curve approaches its limit value for an ideal gas; in particular via eqn.~\ref{omega_id} we get $\lim_{\beta\to0} \xi_2(\beta, \eta) = 1/2$. We can see a steady increase of the timescale by both fixing $\eta$ wile increasing $\beta$ and viceversa. We can interpret the result as follows: the strength of the memory effects increases either by decreasing the temperature or by increasing the depth of the potential well in Fig. \ref{pot}. Both cases correspond to a higher localization of the system in the neighborhood of the lowest minimum of the energy landscape.  

\section{Conclusions}

In this work we presented a coarse-grained analysis of the FPU Model at equilibrium, based on the numerical and analytical study of a generalization of the intermediate scattering function (ISF). Based on MD simulations it is possible to distinguish a \textcolor{red}{crossover} from an ideal gas regime to the integrable limit by progressively cooling the system in the initial state. We then treated analytically the sequence of the Taylor coefficients of the ISF for a wide range of temperatures. Moreover, we presented an analysis of the memory effects in the system, based on the reconstruction of the Generalized Langevin equation which governs the dynamics of the ISF. A quantitative description of the time extent of the memory is possible for a wide range of system parameters. The analysis supports the interpretation of the behavior encountered as a temperature-driven localization on phase space. \\
The study we presented involves several conceptual tools which are largely used in the framework of Liquid State Theory, and applied them to a model traditionally studied in the context of Dynamical Systems. The methods we empolyed can be in general extended to any observable in the system, as e.g.~the normal mode energies or other relevant variables. This would allow to merge the present discussion with the problem of anomalous transport of energy between modes or the propagation of solitary waves, for different initial conditions. The hope of the authors is that, by joining the different approaches some light can be shed on the intriguing and rich phenomenology exhibited by an apparently simple system still able to question the foundations of Statistical Mechanics.

\begin{appendices}
\section{Appendix}
\subsection{Two body potential}\label{app_pot}
In this section we derive the expression for the two-well interaction potential between nearest neighbors  used (eqn.~\ref{pot}). We would like to fix a parametrization in terms of a global energy scale $\alpha$, a length scale $\sigma$ and a factor controlling the unbalance between two wells. Therefore we can define
\begin{equation*}
V(r) = \alpha \Le[\Le(\f r\sigma\Ri)^2 + A\Le(\f r\sigma\Ri)^3 + B \Le(\f r\sigma\Ri)^4\Ri]
\end{equation*}
We then require that the distance $r=\sigma$ corresponds to an equilibrium position, and we impose $V(\sigma)\equiv -\epsilon$, $\epsilon \in \mathbb R$:
\begin{align*}
\f {\mr d V(r)}{\mr d r}\Bigr|_{r=\sigma} &= 0 = \f\alpha\sigma \Le[2+3A+4B\Ri]\\
V(\sigma) &= -\epsilon = \alpha[1 +A+B] \\
\end{align*}
The relations are solved by
\begin{align*}
A &= A(\eta) = -2-4\eta \\
B &= B(\eta) = 1+3\eta 
\end{align*}
With the dimensionless parameter $\eta= \f \epsilon\alpha$ assuming positive values in case the minimum in $r=\sigma$ is negative.
\subsection{Analytical expression for $\omega_n$}\label{omega_n_an}
In this appendix we present the analytical expressions of the series coefficients $\omega_n$; while an exact solution is easily determined at any order in the ideal gas regime, a general expression is not available in case a nonlinear interaction is added to the dynamics. For the non-integrable regime, we can still derive the analytical solution of the lowest orders; in particular in \ref{app_sn_dyn_corr} we compute exactly $\omega_4$, which is the first coefficient that depends on the dynamics.
\label{app_dyn_corr}
\subsubsection{Ideal gas regime}\label{app_free_dyn_corr}
 By expanding the kinetic part of the Liouvillian (eqn.~\ref{liou}) in the general expression in eqn.~\ref{omega_def} we get:
\begin{align}\label{omega_id}
\omega_n^{\free} &=\lla \Le[\Le(\f 1{m}\sum_{\gamma=1}^{N-2}(p_\gamma-p_{\gamma-1})\f{\p}{\p r_\gamma}\Ri)^ne^{iKr_j}\Ri] e^{-iKr_j} \rra_\beta = \Le(\f{iK}{m}\Ri)^n\lla(p_j-p_{j-1})^n\rra_\beta =\Le(\f{iK}{m}\Ri)^n\sum_{k=0}^n\binom{n}{k} \lla p_j^k \rra_\beta\lla (-p_{j-1})^{n-k}\rra_\beta = \nn \\
&\simeq\f 1{\pi} \Le(\f{\beta}{2m}\Ri) \Le(\f{iK}{m}\Ri)^n  \sum_{k=0}^n\binom{n}{k}(-1)^{n-k} \Le[ \int_{-\infty}^{+\infty} \mr d p_j\; e^{-\beta\f{p_j^2}{2m}} p_j^k \Ri] \Le[ \int_{-\infty}^{+\infty} \mr d p_{j-1}\; e^{-\beta\f{p_{j-1}^2}{2m}} p_{j-1}^{n-k} \Ri] = \nn \\
&= \f1{\pi}\Le(\f{\beta}{2m}\Ri)\Le(\f{iK}{m}\Ri)^n\f{(-1)^n+1}{2}\sum_{k=0}^{\f n2}\binom{n}{2k} \Le(\f{2m}\beta\Ri)^{ k+\f 12} \Le[  \int_{-\infty}^{+\infty} \mr d x\; e^{-x^2} x^{2k} \Ri]\times \nn \\
&\times \Le(\f{2m}\beta\Ri)^{\f{n+1}2-k}\Le[  \int_{-\infty}^{+\infty} \mr d x\; e^{-x^2} x^{n-2k} \Ri] = \nn \\
&=\f1{\pi} \Le(\f{\beta}{2m}\Ri)\Le(\f{iK}{m}\Ri)^n \f{(-1)^n+1}{2} \Le(\f{2m}\beta\Ri)^{\f n2+1}\Le[\sum_{k=0}^{\f n2}\binom{n}{2k}  \Gamma\Le(k+\f 12\Ri)\Gamma\Le(\f n2-k+\f 12\Ri) \Ri] = \nn \\
&=\f 1{\pi}(-1)^{\f n2}\Le( \f{2K^2}{m\beta}\Ri)^{\f n2} \f{(-1)^n+1}{2}\sqrt{\pi}2^{\f n2}\Gamma\Le(\f{n+1}2\Ri)=\f 1{\sqrt\pi}(-1)^{\f n2}\Le( \f{4K^2}{m\beta}\Ri)^{\f n2} \f{(-1)^n+1}{2}\Gamma\Le(\f{n+1}2\Ri)
\end{align}
where we used 
\begin{equation*}
\lla p_j^m \rra_\beta = \Le(\f{2m}{\beta}\Ri)^{\f m2}\f 1{\sqrt \pi}\f{(-1)^m+1}2\Gamma\Le(\f {m+1}2\Ri)
\end{equation*}
and
\begin{align*}
&\sum_{k=0}^{\f n2} \binom{n}{2k}\Gamma\left(k+\frac{1}{2}\right)\Gamma\left(\f n2-k+\frac{1}{2}\right)=\sum_{k=0}^{\f n2} \frac{n!}{(2k)!(n-2k)!}\cdot\sqrt{\pi}\,\frac{(2k)!}{4^k k!}\cdot\sqrt{\pi}\,
\frac{(n-2k)!}{4^{\f n2-k} \Le(\f n2-k\Ri)!}\\
&=\frac{\pi n!}{4^{\f n2}\Le(\f n2\Ri)!}\sum_{k=0}^{\f n2} \f {\Le(\f n2\Ri)!}{k!\Le(\f n2 -k\Ri)!} = \frac{\pi n!}{4^{\f n2}\Le(\f n2\Ri)!} \sum_{k=0}^{\f n2} \binom{\f n2}{k}=\frac{\pi n!}{4^{\f n2}\Le(\f n2\Ri)!} 2^{\f n2} =\sqrt \pi 2^{\f n2} \f{\sqrt \pi n!}{4^{n/2}\Le(\f n2\Ri)!} = \sqrt{\pi}2^{\f n2}\Gamma\Le(\f {n+1}{2}\Ri)
\end{align*} 
\subsubsection{$\omega_4$ in the anharmonic regime}\label{app_sn_dyn_corr}
Let us decompose the Liouville operator into kinetic and potential part
\begin{align*}
\mc L = \mc L_r& + \mc L_p \\
\mc L_r\equiv \sum_{j=1}^{N-1}\f 1{m}\Le(p_j- p_{j-1}\Ri)\f{\p}{\p r_j}\;\;\;&\;\;\;\mc L_p\equiv \sum_{j=1}^{N-1}-\f{\p V(r_j)}{\p r_j}\Le(\f{\p }{\p p_j}-\f{\p}{\p p_{j-1}}\Ri) 
\end{align*}
In the following calculation we fix the potential length scale to the value $\sigma=1$ effectively used in the work. The fourth dynamical correlator is then given by:
\begin{align*}
&\omega_4  =(-1)^2\lla \llv (i\mc L)^2 e^{iK r_j}\rrv^2 \rra_\beta = \lla \llv (i\mc L_r+i\mc L_p)^2 e^{i K r_j}\rrv^2\rra_\beta = \lla \llv \Le((i\mc L_r)^2 + (i\mc L_p)^2 + i\mc L_r i\mc L_p + i \mc L_p i\mc L_r\Ri)  e^{i K r_j} \rrv^2 \rra_\beta = \\
& \Bigg\langle \Bigg\vert \sum_{\gamma,\gamma'=1}^{N-2} \Bigg[\f{1}{m^2}(p_\gamma-p_{\gamma-1})(p_{\gamma'}-p_{\gamma'-1})\f{\p^2}{\p r_\gamma r_{\gamma'}} + \f{\alpha}{\sigma m}
\Le(-2\Le(\f{r_\gamma}{\sigma}\Ri)-3A(\eta)\Le(\f{r_\gamma}{\sigma}\Ri)^2-4B(\eta)\Le(\f{r_\gamma}{\sigma}\Ri)^3\Ri)\times\\
&\times\Le(\f{\p}{\p p_\gamma}-\f{\p}{\p p_{\gamma-1}}\Ri)(p_{\gamma'}-p_{\gamma'-1})\f{\p}{\p r_{\gamma'}}\Bigg]e^{iKr_j}\Bigg\vert^2  \Bigg\rangle_\beta = \\
&= \Bigg\langle \Bigg\vert \Le(\f{iK}{m}\Ri)^2 (p_j-p_{j-1})^2  + \f{iK\alpha}{\sigma m}\sum_{\gamma,\gamma'=1}^{N-2}\Le(-2\Le(\f{r_\gamma}{\sigma}\Ri)-3A(\eta)\Le(\f{r_\gamma}{\sigma}\Ri)^2-4B(\eta)\Le(\f{r_\gamma}{\sigma}\Ri)^3\Ri)\times\\
&\times\Le(\delta_{\gamma, \gamma'}-\delta_{\gamma, \gamma'-1}-\delta_{\gamma-1,\gamma'}+\delta_{\gamma-1,\gamma'-1}\Ri)\delta_{\gamma', j} \Bigg\vert^2 \Bigg\rangle_\beta = \\
&= \lla \llv \Le(\f{iK}{m}\Ri)^2 (p_j-p_{j-1})^2  + \f{iK\alpha}{\sigma m}\sum_{\gamma=1}^{N-2}\Le(-2\Le(\f{r_\gamma}{\sigma}\Ri)-3A(\eta)\Le(\f{r_\gamma}{\sigma}\Ri)^2-4B(\eta)\Le(\f{r_\gamma}{\sigma}\Ri)^3\Ri)(2\delta_{\gamma,j}-\delta_{\gamma, j-1}-\delta_{\gamma, j+1})\rrv^2\rra_\beta=\\
&= \Bigg\langle \Bigg\lvert \Le(\f{iK}{m}\Ri)^2 (p_j-p_{j-1})^2  + \f{iK\alpha}{\sigma m}\Big[2\Le(-2\Le(\f{r_j}\sigma\Ri)-3A(\eta)\Le(\f{r_j}\sigma\Ri)^2-4B(\eta)\Le(\f{r_j}\sigma\Ri)^3\Ri)+\\
&-\Le(-2\Le(\f{r_{j-1}}\sigma\Ri)-3A(\eta)\Le(\f{r_{j-1}}\sigma\Ri)^2-4B(\eta)\Le(\f{r_{j-1}}\sigma\Ri)^3\Ri)-\Le(-2\Le(\f{r_{j+1}}\sigma\Ri)-3A(\eta)\Le(\f{r_{j+1}}\sigma\Ri)^2-4B(\eta)\Le(\f{r_{j+1}}\sigma\Ri)^3\Ri)\Big]\Bigg\rvert^2\Bigg\rangle_\beta= \\
&\simeq \Le(\f{K}{m}\Ri)^4  \Big \langle (p_j-p_{j-1})^4 \Big \rangle_\beta + \Le(\f{K\alpha}{\sigma m}\Ri)^2 \Bigg[6\lla \Le(-2\f r\sigma-3A(\eta)\Le(\f r\sigma\Ri)^2-4B(\eta)\Le(\f r\sigma\Ri)^3\Ri)^2\rra_\beta+  \\
&-6  \lla-2\f r\sigma-3A(\eta)\Le(\f r\sigma\Ri)^2-4B(\eta)\Le(\f r\sigma\Ri)^3\rra_\beta^2 \Bigg]  = \\
&= \Le(\f{K}{m}\Ri)^4  \Big \langle (p_j-p_{j-1})^4 \Big \rangle_\beta + 6 \Le(\f{K}{ m}\Ri)^2 \sigma^2(F_\eta, F_\eta) \simeq \Le(\f{K}{m}\Ri)^4 \Le[2\lla p^4\rra_\beta + 6 \lla p^2\rra_\beta^2 \Ri] + 6\Le(\f{K}{m}\Ri)^2\lla F_\eta^2 \rra_\beta 
\end{align*}
where we defined the force acting on a single d.o.f. with
\begin{equation*}
F_{\eta}(r) = -\f{\mr d V_{\eta}(r)}{\mr dr}
\end{equation*}
and the equilibrium variance $\sigma_\beta^2(A, B) = \lla AB\rra_\beta-\lla A \rra_\beta \lla B \rra_\beta$. In the last two lines we reduced the phase averages to one-dimensional integrals, in the assumption of weak statistical correlation between positions and momenta, as discussed in the text. Finally in the last line we noticed that $\sigma^2(F_\eta, F_\eta) = \lla F_\eta^2  \rra_\beta$ because $\lla F_\eta \rra_\beta=0$.

\subsection{Harmonic Limit of the ISF}\label{harm_lim}	
In the limit $\beta \gg 1$ the dynamics of the system approaches the one of a harmonic chain, because the interaction can be effectively be linearized at the bottom of the lowest potential well. In this section we will derive an approximated expression for the ISF in this limit; we will mainly follow the arguments in \cite{yoshida},\cite{shobu} and \cite{radons}. 
The two body potential in this regime can be effectively approximated by its series expansion at order two:
\begin{equation*}
V^h(r_j) = V(\sigma) + \f m2 \Omega^2(q_j-q_{j-1}+\sigma)^2 
\end{equation*}
\begin{equation*}
\Omega^2 \equiv \f{V''(\sigma)}m
\end{equation*}
The linearized dynamics is diagonalized by the \textit{normal modes}, a set of $N-2$ fictitious oscillators parametrized by the \textit{Lagrangian coordinates} $\Le\{\eta_j, \dot \eta_j\Ri\}_{j=1}^{N-2}$; their explicit expression is determined as the discrete Fourier transform of the configurations and momenta; in particular
\begin{align}
q_j(t) &=\sqrt{\f{2}{N-1}}\sum_{k=1}^{N-2} \eta_k(t)\sin\Le(\f {\pi jk}{N-1}\Ri)+j\sigma \label{exp_harm_qj} \\
p_j(t) &=m\sqrt{\f{2}{N-1}}\sum_{k=1}^{N-2} \dot\eta_k(t)\sin\Le(\f {\pi jk}{N-1}\Ri)
\end{align}
The frequency of oscillation of each mode is given by
\begin{equation*}
\omega_j= 2\Omega\sin\Le(\f{\pi j}{2(N-1)}\Ri)
\end{equation*}
We can insert the modes' decomposition of the configurations in the ISF in the quasi-integrable regime $\beta \gg 1$; in this case the phase averages $\lla \cdot \rra_{h, \beta}$ are computed w.r.t. the harmonic Hamiltonian 
\begin{equation*}
H^h(\mb\Gamma) \equiv \sum_{j=1}^{N-2}\f{p_j^2}{2m} + (N-1)V(\sigma) + \f{V''(\sigma)}{2}\sum_{j=1}^{N-1}(q_j-q_{j-1}+\sigma)^2
\end{equation*}
The ISF is then expanded in the following chain of identities:
\begin{align}
&C^h(t) \equiv \lla e^{iK\Le(r_j(t)-r_j(0)\Ri)}\rra_\beta= \lla e^{iK\Le(q_j(t)-q_{j-1}(t)-q_j(0)+q_{j-1}(0)\Ri)} \rra_\beta= \nn\\
&=\lla \exp\Le(iK\sqrt{\f 2 {N-1}}\sum_{l=1}^{N-2}\Le(\eta_l(t)-\eta_l(0)\Ri)\Le[\sin\Le(\f{\pi j l}{N-1}\Ri)-\sin\Le(\f{\pi(j-1)l}{N-1}\Ri)\Ri]\Ri)\rra_\beta= \nn\\
&=\f 1 {Z_\beta}\int_{\mathbb R^{2(N-2)}}\prod_{l=1}^{N-2}\mr d \eta_l\mr d \dot\eta_l\;\exp\Bigg(\sum_{l=1}^{N-2}\Bigg[-\f {m\beta} 2\dot\eta_l^2-m\beta\f{\omega_l^2}2\eta_l^2+iK\sqrt{\f 2 {N-1}}\Le((\cos(\omega_l t)-1)\eta_l(0)+\f{\dot\eta_l(0)}{\omega_l}\sin\Le(\omega_l t\Ri)\Ri)\times \nn\\
&\times\Le(\sin\Le(\f{\pi j l}{N-1}\Ri)-\sin\Le(\f{\pi(j-1)l}{N-1}\Ri)\Ri)\Bigg]\Bigg)= \label{harm_sol}\\
&=\Le(\f{ m\beta}{2\pi}\Ri)^{N-2}\Le(\prod_{m=1}^{N-2}\omega_m\Ri)\Le\{\int_{\mathbb R^{N-2}} \prod_{l=1}^{N-2}\mr d \dot\eta_l\;
\exp\Le(-\f{m\beta}{2}\dot\eta_l^2+iK\sqrt{\f 2 {N-1}}\f{\dot{\eta_l}}{\omega_l}\sin\Le(\omega_l t\Ri)2\cos\Le(\f{\pi(2j-1)l}{N-1}\Ri)\sin\Le(\f{\pi l}{N-1}\Ri)\Ri)\Ri\}\times \nn\\
&\times\Le\{\int_{\mathbb R^{N-2}}\prod_{l'=1}^{N-2}\mr d \eta_{l'}\;\exp\Le(-\f{m\beta\omega_{l'}^2}{2}\eta_{l'}^2+iK\sqrt{\f 2 {N-1}}\eta_{l'}\Le(1-\cos\Le(\omega_{l'} t\Ri)\Ri)2\cos\Le(\f{\pi(2j-1)l'}{N-1}\Ri)\sin\Le(\f{\pi l'}{N-1}\Ri) \Ri)\Ri\}= \label{trigo}\\
&=\Le(\f{ m\beta}{2\pi}\Ri)^{N-2}\Le(\prod_{m=1}^{N-2}\omega_m\Ri)\Bigg\{\int_{\mathbb R^{N-2}}\prod_{l=1}^{N-2}\mr d\dot \eta_l\;\exp\Le(-\Le[\sqrt{\f{m\beta} 2}\dot\eta_l-\f {iK} {\sqrt{m\beta(N-1)}}\f{\sin\Le(\omega_l t\Ri)}{\omega_l}2\cos\Le(\f{\pi(2j-1)l}{N-1}\Ri)\sin\Le(\f{\pi l}{N-1}\Ri) \Ri]^2\Ri)\Bigg\}\times \nn\\
&\times \exp\Le(-\f{4 K^2}{m\beta(N-1)}\f{\sin^2\Le(\omega_l t\Ri)}{\omega_l^2}\cos^2\Le(\f{\pi(2j-1)l}{N-1}\Ri)\sin^2\Le(\f{\pi l}{N-1}\Ri)\Ri) \Bigg\}\times \nn
\end{align}
\begin{align}
&\times\Le\{\int_{\mathbb R^{N-2}}\prod_{l'=1}^{N-2}\mr d \eta_{l'}\;\exp\Le(-\Le[\sqrt{\f{m\beta} 2}\omega_{l'}\eta_{l'}-\f {iK} {\omega_{l'}\sqrt{m\beta(N-1)}}\Le(1-\cos\Le(\omega_{l'} t\Ri)\Ri)2\cos\Le(\f{\pi(2j-1)l'}{N-1}\Ri)\sin\Le(\f{\pi l'}{N-1}\Ri) \Ri]^2\Ri)\Ri\}\times \nn\\
&\times \exp\Le(-\f{4 K^2}{m\omega_{l'}^2\beta(N-1)}\Le(1-\cos\Le(\omega_{l'} t\Ri)\Ri)^2\cos^2\Le(\f{\pi(2j-1)l'}{N-1}\Ri)\sin^2\Le(\f{\pi l'}{N-1}\Ri)\Ri)= \nn \\
&=\Le(\f{m\beta}{2\pi}\Ri)^{N-2}\Le(\prod_{m=1}^{N-2}\omega_m\Ri)\Le(\f{2\pi}{ m\beta}\Ri)^{\f N 2-1}\exp\Le(-\f{4 K^2}{m\beta(N-1)}\sum_{l=1}^{N-2}\f{\sin^2\Le(\omega_l t\Ri)}{\omega_l^2}\cos^2\Le(\f{\pi(2j-1)l}{N-1}\Ri)\sin^2\Le(\f{\pi l}{N-1}\Ri)\Ri)\times \nn\\
&\times\Le(\f{2\pi}{ \beta}\Ri)^{\f N 2-1}\Le(\prod_{m'=1}^{N-2}\omega_{m'}\Ri)^{-1} \exp\Le(-\f{4 K^2}{m\beta(N-1)}\sum_{l'=1}^{N-2}\f{\Le(1-\cos\Le(\omega_{l'} t\Ri)\Ri)^2}{\omega_{l'}^2}\cos^2\Le(\f{\pi(2j-1)l'}{N-1}\Ri)\sin^2\Le(\f{\pi l'}{N-1}\Ri)\Ri)= \nn\\
&=\exp\Le(-\f{8 K^2}{m\beta(N-1)}\sum_{l=1}^{N-2}\f{1-\cos\Le(\omega_l t\Ri)}{\omega_l^2} \cos^2\Le(\f{\pi(2j-1)l}{N-1}\Ri)\sin^2\Le(\f{\pi l}{N-1}\Ri)\Ri) \label{Chh}
\end{align} 
In eqn.~\ref{harm_sol} we inserted the solution for the Lagrangian configurations $\eta_j(t)=\eta_j(0)\cos(\omega_j t)+\f{\dot{\eta}_j(0)}{\omega_j}\sin(\omega_j t)$; in eqn.~\ref{trigo} we inserted the trigonometric identity $\sin\alpha-\sin\beta=2\cos\Le(\f{\alpha+\beta}2\Ri)\sin\Le(\f{\alpha-\beta}2\Ri)$. \\
We can simplify eqn.~\ref{Chh} by approximating the sum in the exponent with its continuum counterpart, according to: 
\begin{align*}
\f{\pi l}{N-1}&\rightarrow x \\
\f{\pi }{N-1}&\rightarrow \mr d x \\
\f{\pi}{N-1}\sum_{l=1}^{N-2} &\rightarrow\int_0^\pi \mr d x
\end{align*}
We then get:
\begin{align}
&C^h(t)\simeq \exp \Le\{-\f{8K^2}{\pi m\beta}\int_0^\pi\mr d x\f{1-\cos(\omega(x) t)}{\omega(x)^2}\cos^2\Le(\f{2j-1}2x\Ri)\sin^2\Le(\f x2\Ri)\Ri\} \label{Cht3} = \\
&=\exp \Le\{-\f{2K^2}{\pi m\beta\Omega^2}\int_0^\pi\mr d x\Le[1-\cos(\omega(x) t)\Ri]\cos^2\Le(\f{2j-1}2x\Ri)\Ri\} \nn = \\
&=\exp \Le\{-\f{2K^2}{\pi m\beta\Omega^2}\Le[\f\pi 2-\int_0^\pi\mr d x\cos(\omega(x) t)\cos^2\Le(\f{2j-1}2x\Ri)\Ri]\Ri\} \label{Cht4}
\end{align}
where we defined
\begin{equation}\label{omega_x}
\omega(x) = 2\Omega\sin\Le(\f x2\Ri)
\end{equation}
The integral term in eqn.~\ref{Cht4} can be rewritten in a more compact fashion via a few additional manipulations; the final expression will also allow us to determine the long-time limit of the harmonic ISF. Via a first integration by part we get:
\begin{align}
&\int_0^\pi \mr d x \cos\Le(\omega(x)t\Ri) \cos^2\Le(\f{2j-1}2 x\Ri)= \int_0^\pi \mr d x \cos\Le(2\Omega\sin\Le(\f x 2\Ri)t\Ri)\cos^2\Le(\f{2j-1}2 x\Ri) = \nn\\
&=\Le[ \cos\Le(2\Omega\sin\Le(\f x 2\Ri)t\Ri)\Le(\f x2+\f{\sin\Le((2j-1)x\Ri)}{2(2j-1)}\Ri)\Ri]_{x=0}^{x=\pi}+\nn\\
&+\int_{0}^\pi \mr d x2\Omega t \sin\Le(2\Omega t \sin\Le(\f x 2\Ri)\Ri) \f 12 \cos\Le(\f x 2\Ri)\Le[\f x2+\f{\sin\Le((2j-1)x\Ri)}{2(2j-1)}\Ri]= \nn\\
&\simeq \f \pi 2\cos\Le(2\Omega t\Ri)+\f{\Omega t}2\int_0^\pi \mr d x \sin\Le(2\Omega t \sin\Le(\f x 2\Ri)\Ri)\cos\Le(\f x 2\Ri) x \label{bessel1}
\end{align}
where we noticed that the contribution of the last term is negligible for $j=N/2\gg 1$ Via an additional integration by parts in the second term of eqn.~\ref{bessel1} we get:
\begin{align}
&\int_0^\pi \mr d x \sin\Le(2\Omega t \sin\Le(\f x 2\Ri)\Ri)\cos\Le(\f x 2\Ri) x = -\Le[ x \cos\Le(2\Omega t\sin\Le(\f x 2\Ri)\Ri)\f 1 {\Omega t}\Ri]_{x=0}^{x=\pi}+\int_{0}^{\pi} \mr d x \cos\Le(2\Omega t\sin\Le(\f x 2\Ri)\Ri)\f 1 {\Omega t} = \nn \\
&=\f \pi {\Omega t}\Le[J_0(2\Omega t)-\cos\Le(2\Omega t\Ri)\Ri]
\end{align}
We finally obtain the following analytical approximation for the ISF in the harmonic regime:
\begin{equation}\label{Cht6}
C^h(t) = \exp\Le[-\f{K^2}{m \beta\Omega^2}\Le( 1 -J_0(2\Omega t)\Ri)\Ri]
\end{equation}
which is normalized such that $C^h(0)=1$. The oscillating pattern of the function is encapsuled in the periodic behavior of the Bessel function $J_0(\Omega t)$. Within this approximation it is straightforward to determine
\begin{equation}
\lim_{t\to+\infty} C^h(t) = e^{-\f{K^2}{m\beta\Omega^2}} \label{Cht5}
\end{equation}

\subsection{$\omega_n$ in the harmonic regime}
In this Appendix we present an exact method for the numerical calculation of the Taylor coefficients of the ISF in the harmonic limit. The series expansion of the correlation function in this regime can be directly computed from the time derivatives of eqn.~\ref{Cht6}:
\begin{equation}\label{om_harm1}
\omega_{2n}^h=\Le.\f{\mr d^{2n} }{\mr d t^{2n}} \exp\Le(-\f{K^2}{m\beta\Omega^2}(1-J_0(2\Omega t))\Ri)\Ri|_{t=0}
\end{equation}
In order to compute these derivatives, we can make use of Faa' di Bruno's formula for the $2n$-th derivative of a composite function:
	\begin{align}
	\f{\mr d^{2n}}{\mr d x^{2n}}f\Le(g(x)\Ri) &= \sum_{\mb m\in \mc K_{2n,2n}} \f{(2n)!}{m_1!m_2!\cdots m_{2n}!}f^{(m_1+\cdots+m_{2n})}\Le(g(x)\Ri)\prod_{j=1}^{2n}\Le(\f{g^{(j)}(x)}{j!}\Ri)^{m_j} \label{faa1}\\
	\mc K_{n,s} &\equiv \Le\{ \mb m\in \mathbb N_0^{n}: \; \;\sum_{l=0}^{n} l\cdot m_l = s \Ri\} \label{faa2}
	\end{align}
	by establishing the correspondences
	\begin{align*}
	x &\lrar t \\
	f(x) &\lrar \exp(t) \\
	g(x) &\lrar -\frac{K^2}{m\beta\Omega^2}(1-J_0(2\Omega t))
	\end{align*}
	We now need to express the different contributions appearing in eqn.~\ref{faa1} in terms of the functions we are interested in. We note
	\begin{equation}\label{fms}
	f^{(m_1+\cdots+m_{2n})}\Le(g(x)\Ri) \lrar \exp\Le(-\frac{K^2}{m\beta\Omega^2}(1-J_0(2\Omega t))\Ri)
	\end{equation} 
	and
	\begin{equation}\label{diffJ0}
	g^{(2j)}(x) \lrar \f{\mr d^{2j}}{\mr d t^{2j}} \Le[-\f{K^2}{m\beta\Omega^2}(1-J_0(2\Omega t))\Ri] = \f{K^2}{m\beta\Omega^2}\f{\mr d^{2j}}{\mr d t^{2j}}J_0(2\Omega t)
	\end{equation}
	where we omitted the constant term as in eqn.~\ref{faa1} we are only interested in strictly positive derivatives ($j\ge 1$). We can get a closed expression for the derivatives of $J_0(2\Omega t)$ as follows:
	\begin{align}\label{expJ0}
	J_0(2\Omega t')=\Le.J_0\Le(2\Omega(t+t')\Ri)\Ri|_{t=0}=\exp\Le.\Le(2\Omega t'\f{\mr d}{\mr d (2\Omega t)}\Ri)J_0(2\Omega t)\Ri|_{t=0}=\sum_{j=0}^{+\infty}\f{{t'}^j}{j!}\Le.\f{\mr d^j}{\mr d t^j}J_0(2\Omega t)\Ri|_{t=0}=\sum_{j=0}^{+\infty}\f{(-1/4)^j}{j!^2}(2\Omega t')^{2j} 
	\end{align}
	where the last identity stems from the Taylor expansion of $J_0(2\Omega t')$. 
	We can then equate term by term the same powers of $t'$ in the last identity of eqn.~\ref{expJ0} to get $ \forall{ j \ge 0}$:
	\begin{align*}
	\Le.\f{\mr d^{2j}}{\mr d t^{2j}}J_0(2\Omega t)\Ri|_{t=0} &= (-1)^j\f{(2j)!}{j!^2}\Omega^{2j}=(-1)^j\binom{2j}{j}\Omega^{2j}
	\end{align*}
	while the odd derivatives are identically null, as $J_0$ is an even function. This in particular implies $m_{2j+1}\equiv 0$  $\forall j\ge 0$ in eqn.~\ref{faa1}. Due to this symmetry, the set in eqn.~\ref{faa2} can be rewritten as
	\begin{align*}
	&\tilde{\mc K}_{2n,s} \equiv \Le\{ \mb m\in \mathbb N_0^{2n}: \; \;\sum_{l=0}^{n}2 l\cdot m_{2l} = s, \;\;m_{2l+1}\equiv 0 \;\;\forall \;\;0\le l\le n-1\Ri\}
	\end{align*} 
	such that
	\begin{align}
	\omega_{2n}^h=&\Le.\f{\mr d^{2n} }{\mr d t^{2n}} \exp\Le(-\f{K^2}{m\beta\Omega^2}(1-J_0(2\Omega t))\Ri)\Ri|_{t=0} = \sum_{\mb m\in \tilde{\mc K}_{2n,2n}} \f{(2n)!}{m_2!m_4!\cdots m_{2n}!}\prod_{j=1}^{n}\Le(\f{(-1)^jK^2}{m\beta\Omega^2(2j)!}\binom{2j}{j}\Omega^{2j}\Ri)^{m_{2j}} =\nn \\
	&=  \sum_{\mb m\in \tilde{\mc K}_{2n,2n}} \f{(2n)!}{m_2!m_4!\cdots m_{2n}!}\Le(-\Omega^2\Ri)^{\sum_{j=1}^n j m_{2j}}\Le(\f{K^2}{m\beta\Omega^2}\Ri)^{\sum_{j=1}^n m_{2j}}\prod_{j=1}^n\Le(\f 1 {j!}\Ri)^{2m_{2j}}= \nn \\
	&= \Le(-\Omega^2\Ri)^{n}\sum_{\mb m\in \tilde{\mc K}_{2n,2n}} \f{(2n)!}{f(\mb m)}\Le(\f{K^2}{m\beta\Omega^2}\Ri)^{g(\mb m)}h(\mb m) \label{d2_exp}
	\end{align}
	where we defined
	\begin{align*}
	f(\mb m)\equiv \prod_{j=1}^n m_{2j}! \hspace{7mm} g(\mb m)\equiv\sum_{j=1}^n m_{2j}  \hspace{7mm} h(\mb m)\equiv\prod_{j=1}^n\Le(\f 1 {j!}\Ri)^{2m_{2j}}
	\end{align*}
	The elements of $\tilde K_{2n, 2n}$ can be determined in a recursive routine as described in \cite{maiocchi}. The values in eqn.~\ref{d2_exp} can then be computed at arbitrarily high orders.

\subsection{Liouvillian in non canonical coordinates}\label{change_liou}
We derive here a symmetric expression for the Liouvillian in the set of non canonical coordinates $(\mb r,\mb p)$ used in the work.
\begin{align*}
i\mc L &=\sum_{i=1}^{N-2}\Le[-\f{\p V(\mb q)}{\p q_i}\f{\p}{\p p_i}+\f{p_j}{m}\f{\p}{\p q_i}\Ri] = \sum_{i, j=1}^{N-2}\Le[-\f{\p r_j}{\p q_i}\f{\p V(\mb q)}{\p r_j}\f{\p}{\p p_i}+\f{p_j}{m}\f{\p r_j}{\p q_i}\f{\p}{\p r_k}\Ri] \\ 
&= \sum_{i, j =1}^{N-2} \Le[ -\f{\p (q_j-q_{j-1})}{\p q_i}\f{\p V(\mb q)}{\p r_j}\f{\p}{\p p_i}+\f{p_i}{m}\f{\p (q_j-q_{j-1})}{\p q_i}\f{\p}{\p r_j} \Ri] = \\
&= \sum_{i, j =1}^{N-2} \Le[ -(\delta_{i,j}-\delta_{i,j-1})\f{\p V(\mb q)}{\p r_j}\f{\p}{\p p_i}+\f{p_i}{m}(\delta_{i, j}-\delta_{i,j-1})\f{\p}{\p r_j} \Ri] = \\
&= \sum_{i=1}^{N-2} \Le[ -\Le(\f{\p V(\mb q)}{\p r_i}-\f{\p V(\mb q)}{\p r_{i+1}}\Ri)\f \p {\p p_i}+\f{p_i}{m}\Le(\f{\p}{\p r_i}-\f{\p}{\p r_{i+1}}\Ri)\Ri] = \\
&= - \Le(\f{\p V(\mb q)}{\p r_1}-\f{\p V(\mb q)}{\p r_2}\Ri) \f{\p}{\p p_1}+\f{p_1}{m} \Le(\f{\p}{\p r_1}-\f{\p}{\p r_2}\Ri)  -\Le(\f{\p V(\mb q)}{\p r_2}-\f{\p V(\mb q)}{\p r_3}\Ri)\f{\p}{\p p_2}+\f{p_2}{m} \Le(\f{\p}{\p r_2}-\f{\p}{\p r_3}\Ri)+\cdots+\\
&- \Le(\f{\p V(\mb q)}{\p r_{N-2}}-\f{\p V(\mb q)}{\p r_{N-1}}\Ri) \f{\p}{\p p_{N-2}}+\f{p_{N-2}}{m} \Le(\f{\p}{\p r_{N-2}}-\f{\p}{\p r_{N-1}}\Ri) =
\end{align*}
\begin{align*}
&=-\f{\p V(\mb q)}{\p r_1}\Le( \f{\p}{\p p_1} - \boxed{\f{\p}{\p p_0}} \Ri) + \f 1{m}(p_1-\boxed{p_0})\f{\p }{\p r_1}+\cdots-\f{\p V(\mb q)}{\p r_{N-1}}\Le( \boxed{\f{\p}{\p p_{N-1}}} - \f{\p}{\p p_{N-2}} \Ri) + \f 1{m}(\boxed{p_{N-1}}-p_{N-2})\f{\p }{\p r_{N-1}}= \\
&=\sum_{i=1}^{N-1} \Le[-\f{\p V(r_i)}{\p r_i}\Le(\f{\p }{\p p_i}-\f{\p}{\p p_{i-1}}\Ri) + \f 1{m}\Le(p_i- p_{i-1}\Ri)\f{\p}{\p r_i} \Ri]
\end{align*}
In the second last line we added to the sum the terms in boxes, as they are identically null due to the boundary conditions considered.
\subsection{Polynomial expansion of the dynamical correlators}\label{app_spread}
\begin{figure}[H]
	\begin{center}
		\includegraphics[scale=0.6]{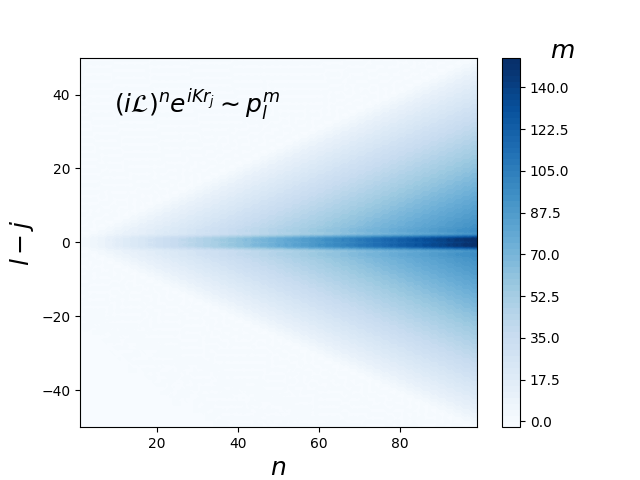}
		\caption{Leading power $m$ for momentum of the $l-$th particle in $(i\mc L)^n e^{iKr_j}$}\label{contour_plot_liou2}
	\end{center}
\end{figure}

In this Appendix we expand the discussion in Section \ref{constr}, in order to gain a deeper insight in the spreading operators. The dynamics of the ISF at finite times can be reconstructed via a recursive calculation of the coefficients of the polynomial expansion in eqn.~\ref{general_exp}. From the expression of $i\mc L$ in eqn.~\ref{liou} we note that only the d.o.f. $j$ and $j-1$ are effectively involved in $i\mc Le^{iKr_j}$. At second order (i.e. in $(i\mc L)^2 e^{iKr_j}$) we have to add an additional layer of neighbors around $j$ and so on. The whole chain is finally expected to participate in the dynamics for $n\gtrsim \floor*{\f N 2}$, while for higher orders additional effects will enter the evolution, due to reflections at the ends of the chain and interference of different wavefronts. In the following we will omit the analysis of these finite size effects; as $N\gg 1$ the dynamics at short times is not affected by the boundaries. \\
A graphical representation of the spreading process is given in Fig. \ref{contour_plot_liou2}, where we plot for different particles $l$ on the left and right of $j$ the leading power $m$ in the momentum at order $n$, such that $(i\mc L)^n\sim p_l^m e^{iKr_j}$. We observe a monotonic extension of the non-locality for increasing values of $n$. 
Once the maximum order of the propagation $n=n_{max}$ has been fixed, it becomes convenient to quantify boundaries of the domain at this given order. This way it is possible to allocate the minimum amount of CPU memory which is needed for the storage of the coefficients of the tensor $\mc I$. \\
We present in the following a detailed calculation of the contributions in the Liouville operator controlling the features described above. We leave the remaining part of the Appendix to the readers interested in the most technical aspects of the analysis. \\ \\
By following the same arguments leading to the definition of eqn.~\ref{sigmarR}, we can define the contribution in $(i\mc L)^n e^{iKr_j}$ which connects the central particle of the chain with the farthest on its left:
\begin{align}
&{\sigma_r^L}(n, j, j-k  ) = \theta\Le(\floor*{\f n2}-j+k\Ri)\Le[ \Le( -\f{\p V}{\p r_k} \Ri)\Le(\f{\p}{\p p_k}-\f{\p}{\p p_{k-1}}\Ri)\Ri]^{\theta(n-2)}\times \nn \\
&\times\Le\{ \prod_{i=0}^{ j-k-2} \Le[ \f 1{m}(p_{j-i-1}-p_{j-i-2}) \f{\p}{\p r_{j - i-1}}\Ri] \Le[ \Le( -\f{\p V}{\p r_{j-i-1}} \Ri)\Le(\f{\p}{\p p_{j-i-1}}-\f{\p}{\p p_{j-i-2}}\Ri)\Ri]\Ri\}^{\theta\Le(n-4\Ri)}\Le[\f 1{m}(p_j-p_{j-1})\f{\p}{\p r_j}\Ri]^{\theta \Le(n-2 \Ri)}  \label{sigmarL}\\
&{\sigma_p^R}(n, j, l-j) = \theta\Le(\floor*{\f{n-1}2}-l+j\Ri) \Le\{ \prod_{i=0}^{l-j-1 } \Le[ \f 1{m}(p_{j+ i+1}-p_{j+i}) \f{\p}{\p r_{j + i+1}}\Ri] \Le[ \Le( -\f{\p V}{\p r_{j + i+1}} \Ri)\Le(\f{\p}{\p p_{j+ i+1}}-\f{\p}{\p p_{j+ i}}\Ri)\Ri]\Ri\} ^{\theta\Le(n-3\Ri)}\times \nn \\
&\times \Le[\f 1{m}(p_j-p_{j-1})\f{\p}{\p r_j}\Ri]^{\theta \Le(n-1 \Ri)} \label{sigmapR} \\
&{\sigma_p^L}(n, j, j-l) = \theta\Le(\floor*{\f{n+1}2}-j+l\Ri) \Bigg\{ \prod_{i=0}^{j-l-2 } \Le[ \f 1{m}(p_{j-i-1}-p_{j-i-2}) \f{\p}{\p r_{j - i-1}}\Ri]\times \nn \\
&\times \Le[ \Le( -\f{\p V}{\p r_{j -i-1}} \Ri)\Le(\f{\p}{\p p_{j-i-1}}-\f{\p}{\p p_{j- i-2}}\Ri)\Ri]\Bigg\} ^{\theta\Le(n-3\Ri)} \Le[\f 1{m}(p_j-p_{j-1})\f{\p}{\p r_j}\Ri]^{\theta \Le(n-1 \Ri)} \label{sigmapL} 
\end{align}
From eqn.~\ref{sigmapR} and eqn.~\ref{sigmapL} we can extract the momenta entering the dynamics at order $n$:
\begin{align}\label{Lp}
L_p(n, j) &\equiv \Le\{j-\floor*{\f{n+1}2},\cdots, j+\floor*{\f{n-1}2} \Ri\}\in \mathbb N^{\Le(\floor*{\f {n+1} 2}+\floor*{\f {n-1} 2}+1\Ri)}  \equiv \Le\{l_{\min},\cdots, l_\max\Ri\}
\end{align}
and we notice a slight asymmetry w.r.t. $j$ as a consequence of the dynamics propagating slightly faster on the left side of $j$, as eqn.~\ref{liou} suggests. 

Another relevant factor tuning the numerical complexity of the problem is the maximum power of a general displacement $r_k$ and momentum $p_l$ in eqn.~\ref{general_exp} at a certain order. This piece of information fixes the dimensions of the indeces $\mb m$ and $\mb s$ of $\mc I$. The calculation of this leading order is easily performed for the d.o.f. adjacent to $j$, as we do not need to account for finite spreading of the dynamics along the chain. Let us focus in particular to the leading power of $r_k$ for $\lv k-j\rv\le 1$; the relevant contribution in $(i\mc L)^n e^{iK r_j}$ is
\begin{equation}\label{lead_rk}
\Le[\Le(-\f{\p V(r_k)}{\p r_k}\Ri)\Le(\f{\p}{\p p_k}-\f{\p}{\p p_{k-1}}\Ri)(p_j-p_{j-1})\f{\p}{\p r_j}\Ri]^{\floor*{\f{n}2}}e^{iK r_j} \sim r_k^{M_r} e^{iK r_j}
\end{equation}
with $M_r=2\Le(\floor{\f n2}-1\Ri)+3 = 2\floor{\f n2}+1$. It is analogously straightforward to identify that the contribution in $(i\mc L)^n e^{iKr_j}$ associated to leading power of $p_l^{s_l}$ for $l\in\{j, j-1\}$ is
\begin{equation}\label{lead_pl}
\Le[\Le(p_j-p_{j-1}\Ri)\f{\p}{\p r_j}\Ri]^ne^{iK r_j} \sim p_l^{S_p}
\end{equation}
with $S_p=n$. For another d.o.f. arbitrarily far from $j$ we could proceed as in eqn.~\ref{sigmapR}-\ref{sigmapL} by defining suitable chains of differential operators in $(i\mc L)^n e^{iKr_j}$. A subset of $\mc N_{r/p}^{(R/L)}<n$ terms must be devoted to the spreading of the interaction till the d.o.f. in scope is reached. Via direct counting in eqn.~\ref{sigmarR}, \ref{sigmarL}, \ref{sigmapR}, \ref{sigmapL} respectively we can determine
\begin{align*}
\mc N_r^R(n, j, k-j)  &= \theta\Le(\floor*{\f n2}-k+j \Ri)\Le[2\theta(n-2)+2(k-j-1)\theta(n-4)\Ri]\\ 
\mc N_r^L(n, j, j-k)  &= \theta\Le(\floor*{\f n2}-j+k \Ri) \Le[2\theta(n-2)+2(j-k-1)\theta(n-4)\Ri]\\
\mc N_p^R(n, j, l-j)  &= \theta\Le(\floor*{\f{n-1}2}-l+j\Ri)  \Le[\theta(n-1)+2(l-j)\theta(n-3)\Ri] \\
\mc N_p^L(n, j, j-l)  &= \theta\Le(\floor*{\f{n+1}2}-j+l\Ri)  \Le[\theta(n-1)+2(j-l-1)\theta(n-3)\Ri] \\
\end{align*}
Once the desired coordinate has been reached, the remaining $n-\mc N_{r/p}^{(R/L)}$ factors must be combined in specific sequences in order to provide the differentials associated to the leading powers in $r_k^{M_r}$ or $p_l^{S_p}$. Such operators are respectively defined as
\begin{align}
\tilde {\sigma}_r^{(R/L)}(n,j, k ) & = \Le\{\Le[-\f{\p V}{\p r_k}\Le(\f{\p}{\p p_k}-\f{\p}{\p p_{k-1}}\Ri)\Ri]\Le[\f 1{m_p}(p_k-p_{k-1})\f{\p }{\p r_k}\Ri]\Ri\} ^{\floor*{\f {n-\mc N_r^{(R/L)}(n,j,\lv k-j\rv)}2}}\sigma_r^{(R/L)}(n, j, \lv k -j\rv) \label{sigmatil1}\\
\tilde {\sigma}_p^{(R/L)}(n,j, l ) & =\Le\{\Le[  \f 1{m_p}(p_l-p_{l-1}) \f{\p}{\p r_l}\Ri]^2 \Le[ -\f{\p V}{\p r_l}\Le(\f{\p}{\p p_l}-\f{\p}{\p p_{l-1}}\Ri)\Ri]\Ri\}^{\mc {N^'}^{(R/L)}_p(n,j,\lv l -j \rv)} \times \nn \\
&\times \Le\{\Le[ \f 1{m_p}(p_l-p_{l-1}) \f{\p}{\p r_l}\Ri]^3 \Le[-\f{\p V}{\p r_l}\Le(\f{\p}{\p p_l}-\f{\p}{\p p_{l-1}}\Ri)\Ri]\Ri\}^{\floor*{\f{n-\mc N_p^{(R/L)}(n,j,\lv l -j \rv)}4}} \sigma_p^{(R/L)}(n, j, \lv l -j \rv) \label{sigmatil2}
\end{align}
with
\begin{equation*}
\mc {N^'}^{(R/L)}_p(n,j, \lv l -j \rv)=\floor*{\f 13\Le(n-{\mc N_p}^{(R/L)}(n,j, \lv l -j \rv)-4 \floor*{\f{n-{\mc N_p}^{(R/L)}(n,j, \lv l -j \rv)}4}\Ri)} \in\{0,1\}
\end{equation*}
Note that eqn.~\ref{sigmatil1} and eqn.~\ref{sigmatil2} are symmetric w.r.t. the index $j$, in contrast to the expression of the spreading operators themselves. We are now almost able to write the explicit expression of the functions $\mc M_r(n,j,\lv k-j\rv)$ and $\mc S_p(n,j,\lv k-j\rv)$, returning respectively the maximum power of $r_k$ and $p_l$ at order $n$. This can be accomplished via direct counting in eqn.~\ref{sigmatil1} and eqn.~\ref{sigmatil2}. We have to distinguish three different cases; for non null contributes we get:
\begin{align*}
1. \floor*{\f {n-\mc N_r^{(R/L)}(n,j,\lv k-j\rv)}2} =0 \implies&  { \mc M}_r(n,j, \lv k-j \lv ) =3 \\
2. \floor*{\f {n-\mc N_r^{(R/L)}(n,j,\lv k-j\rv)}2} =1 \implies&  {\mc M}_r(n,j, \lv k-j \lv ) = 5 \\
3. \floor*{\f {n-\mc N_r^{(R/L)}(n,j,\lv k-j\rv)}2} \ge 2 \implies& {\mc M}_r(n,j, \lv k-j \lv ) = 6 +2\Le(\floor*{\f{n-N_r^{(R/L)}(n,j,\lv k-j\rv)}{2}}-1\Ri)
\end{align*}
and therefore
\begin{align*}
&\mc M_r(n, j, \lv k-j \lv ) =
&=\begin{cases}
2\floor{\f n2}+1, &0 \le  \llv j-k \rrv \le 1 \\
\theta\Le(\floor*{\f n2}-\lv k-j\rv \Ri)\Big\{3+2\theta\Le(\floor*{\f {n-\mc N_r^{(R/L)}(n,j,\lv k-j\rv)}2}-1\Ri)+ \\
+\theta\Le(\floor*{\f{n-\mc N_r^{(R/L)}(n,j,\lv k-j\rv)}2}-2\Ri) \Le(2\floor*{\f{n-\mc N_r^{(R/L)}(n,j,\lv k-j\rv)}2} -1\Ri)\Big\}, &\text{otherwise}
\end{cases}
\end{align*}
Analogously we can write for the momenta:
\begin{align*}
1. \floor*{\f {n-\mc N_p^{(R/L)}(n,j,\lv l-j\rv)}4} =0 \implies& \mc S_p(n,j,\lv l-j \lv) = 1+\theta\Le( {\mc N'}^{(R/L)}_p(n,j,\lv l-j\rv)-1\Ri)\\
2.  \floor*{\f {n-\mc N_p^{(R/L)}(n,j,\lv l-j\rv)}4} =1 \implies& \mc S_p(n,j,\lv l-j \lv) = 3+\theta\Le( {\mc N'}^{(R/L)}_p(n,j,\lv l-j\rv)-1\Ri) \\
3. \floor*{\f {n-\mc N_p^{(R/L)}(n,j,\lv l-j\rv)}4} \ge 2 \implies& \mc S_p(n,j,\lv l-j \lv) = 1+2\Le(\floor*{\f {n-\mc N_p^\pm(n,j,\lv l-j\rv)}4}-1\Ri)+\\&+3\theta\Le( - {\mc N'}^{(R/L)}_p(n,j,\lv l-j\rv) \Ri)+4\theta\Le( {\mc N'}^{(R/L)}_p(n,j,\lv l-j\rv)-1\Ri) 
\end{align*}
The different cases above can be expressed together to get:

\begin{align*}
\mc S_p(n,j,\lv l-j \lv) &= 
\begin{cases}
n,  &l \in \{j, j-1\} \\
\theta\Le(\floor*{\f{n\mp1}{2}}-\lv j-l\rv\Ri)\Big\{ 1 +2\theta\Le(\floor*{\f{n-\mc N_p^{(R/L)}(n,j,\lv l-j\rv)}4}-1\Ri) +&\\ +\theta\Le(\floor*{\f{n-\mc N_p^{(R/L)}(n,j,\lv l-j\rv)}4}-2\Ri)\times &\\
\times \Big[2 \Le(\floor*{\f{n-N_p^{(R/L)}(n,j,\lv l-j\rv)}4}- 2 \Ri)+3\theta\Le(-{\mc N'}^{(R/L)}_p(n,j,\lv l-j\rv)\Ri)+&\\
+3\theta\Le({\mc N'}^{(R/L)}_p(n,j,\lv l-j\rv)-1\Ri)\Big]+\theta\Le({\mc N'}^{(R/L)}_p(n,j,\lv l-j\rv)-1\Ri)\Big\}, &  \text{otherwise} \\
\end{cases}
\end{align*}	

\subsection{Recursive construction of the dynamical tensor}\label{app_recursion}
We complete here the derivation of the recursion relations for the construction of the tensor $\mc I^{(n)}$ defined in eqn.~\ref{general_exp}. The iterative scheme for $i\mc L_p^\gamma$ has been derived in eqn.~\ref{ilpgamma}. We can proceed in the same fashion for the other differential operators in $i\mc L$. Let us now consider the action of the operator
\begin{equation*}
i\mc L_r^\gamma \equiv \f 1 m (p_\gamma-p_{\gamma-1})\f{\p}{\p r_\gamma}, \hspace{5mm}  k_\min \le \gamma \le k_\max
\end{equation*}
on a generic monomial of the expansion at order $n$. We have:
\begin{align*}
&\Le\{ i\mc L_r^\gamma \Le[ r_{k_{\min}}^{m_{k_{\min}}} \cdots r_\gamma^{m_\gamma} \cdots r_{k_{\max}}^{m_{k_{\max}}} \Ri]\Le[ p_{l_{\min}}^{s_{l_{\min}}} \cdots p_{\gamma-1}^{s_{\gamma-1}} p_\gamma^{s_\gamma} \cdots p_{l_{\max}}^{s_{l_{\max}}} \Ri] \Ri\} e^{iKr_j}= \\
=& \f {m_\gamma} m\Le\{ \Le[ r_{k_{\min}}^{m_{k_{\min}}} \cdots  r_\gamma^{m_\gamma-1} \cdots r_{k_{\max}}^{m_{k_{\max}}} \Ri]\Le[ p_{l_{\min}}^{s_{l_{\min}}} \cdots \Le(   p_{\gamma-1}^{s_{\gamma-1}} p_\gamma^{s_\gamma+1}-p_{\gamma-1}^{s_{\gamma-1}+1} p_\gamma^{s_\gamma}\Ri) \cdots p_{l_{\max}}^{s_{l_{\max}}} \Ri] \Ri\}
\end{align*}
We can then infer the following recursion relation:
\begin{equation}\label{ilrgamma1}
{\mc I}^{(n+1)}_{\mb{m,s}}\Big\vert_{r,\gamma} \equiv \f {m_\gamma} m \sum_{k=0,1}(-1)^k \mc I^{(n)}_{\mb{m}-\mb{\hat e}_\gamma,\mb s+\mb{\hat e}_{\gamma-k}}
\end{equation} 
We finally have to consider separately the action on $i\mc L_r^j$ on $e^{iKr_j}$:
\begin{align*}
&\Le\{ \Le[ r_{k_{\min}}^{m_{k_{\min}}} \cdots  r_{k_{\max}}^{m_{k_{\max}}} \Ri]\Le[ p_{l_{\min}}^{s_{l_{\min}}} \cdots p_{j-1}^{s_{j-1}} p_j^{s_j} \cdots p_{l_{\max}}^{s_{l_{\max}}} \Ri] \Ri\}i\mc L_r^j e^{iKr_j} = \\
=&\f {iK} m \Le\{ \Le[ r_{k_{\min}}^{m_{k_{\min}}}\cdots r_{k_{\max}}^{m_{k_{\max}}} \Ri]\Le[ p_{l_{\min}}^{s_{l_{\min}}} \cdots \Le(   p_{j-1}^{s_{j-1}} p_j^{s_j+1}-p_{j-1}^{s_{j-1}+1} p_j^{s_j}\Ri) \cdots p_{l_{\max}}^{s_{l_{\max}}} \Ri] \Ri\} e^{iKr_j}\\
\end{align*}
and therefore
\begin{equation}\label{ilj2}
{\mc I}^{(n+1)}_{\mb{m,s}}\Big\vert_{r,j, K} \equiv \f {iK} m \sum_{k=0,1}(-1)^k \mc I^{(n)}_{\mb{m},\mb s+\mb{\hat e}_{j-k}}
\end{equation}
The final expression for the dynamical tensor at order $n+1$ is obtained by summing the contributions of eqn.~\ref{ilpgamma}, eqn.~\ref{ilrgamma1} and eqn.~\ref{ilj2} over the subset $\bs \Gamma$ of the coordinates entering the interaction at order $n$:
\begin{equation*}
{\mc I}^{(n+1)}_{\mb{m,s}} = \sum_{\gamma\in\bs \Gamma}\Le({\mc I}^{(n+1)}_{\mb{m,s}}\Big\vert_{p,\gamma}+{\mc I}^{(n+1)}_{\mb{m,s}}\Big\vert_{r,\gamma}\Ri)+{\mc I}^{(n+1)}_{\mb{m,s}}\Big\vert_{r,j, K}
\end{equation*}
with the initial condition
\begin{equation*}
\mc I^{(0)}_{\mb m,\mb s}=
\begin{cases}
1, &\text{for}\;\;\mb m= \mb e_j, \mb s=\mb 0 \\
0, & \text{elsewhere}
\end{cases}
\end{equation*}
Let us call $n_\max$ the maximum order of $\mc I^{(n)}$ we aim to compute. 
The set $\bs \Gamma$ can be fixed once for all for any $n\in \{0,\cdots, n_\max\}$ : 
\begin{equation*}
\bs \Gamma \le \{\min\{K_r(n_\max, j), L_p(n_\max, j)\}, \cdots , \max\{K_r(n_\max, j), L_p(n_\max, j)\} \}= \Le\{j-\floor*{\f{n_\max+1}2},\cdots, j+\floor*{\f {n_\max}{2}}\Ri\}
\end{equation*}
\subsection{Analysis of the radius of convergence}\label{conv_radius} 
The series expansion in eqn.~\ref{ck_tay} can be formally written for any $t\in\mathbb C$. However, the domain ensuring a finite convergence of the expansion is in general bounded to compact disks centered in the origin with an extension bounded by a finite radius of convergence $r$. In this appendix we briefly discuss the calculation of $r$ and how it connects to the relaxation of the correlation function. Let us rewrite 
\begin{align*}
C_j(t)&=\sum_{m=0}^{+\infty}c_mt^m\\
c_m&=\begin{cases}
\f{\omega_{m}}{m!}& \text{$m$ even} \\
0 & \text{$m$ odd}
\end{cases}
\end{align*}
The \textit{Cauchy-Hadamard Theorem} relates the radius of convergence $r$ to the series' coefficients according to 
\begin{equation*}
r=\f 1 {\limsup_{n\to+\infty}\sqrt[n]{\llv c_n\rrv}}=\lim_{n\to+\infty}\sqrt[n]{\f{(2n)!}{\llv \omega_{2n}\rrv}}
\end{equation*} 
The calculation of $r$ is therefore possible from the estimate of the limit of the sequence
$\alpha_n\equiv1 /\sqrt[n]{\llv c_n\rrv}$. The fist values of $\alpha_n$ are shown in the following figure:
\begin{figure}[H]
	\begin{center}
		\includegraphics[scale=0.55]{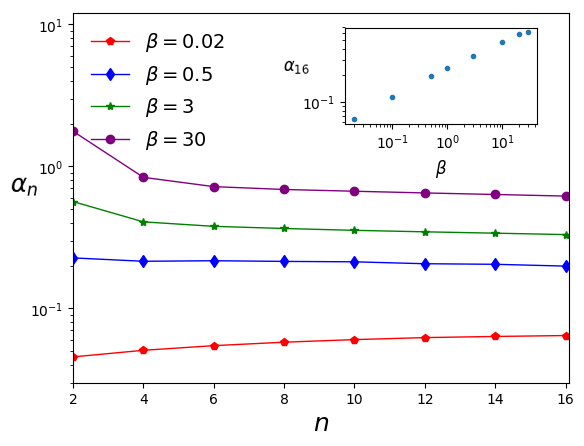}
		\caption{First terms of the sequence $\alpha_n$, for four values of the inverse temperatures $\beta$; inset: estimate of the radius of convergence as a function $\beta$ VIA $\alpha_16$.}
	\end{center}
\end{figure}
We can compute exactly the sequence $\alpha_n$ in the case of an ideal gas, in order to determine the related radius of convergence $r^\free$:
\begin{align*}
\alpha_n^\free &= \f 1 {\sqrt[n]{\llv c_n\rrv}}=
\sqrt[n]{\f{(2n)!}{\llv \omega_{2n}^{\free}\rrv}}=
\sqrt[n]{\f{(2n)!}{\Le(\f{4K^2}{m\beta}\Ri)^n\f{(2n)!}{4^n n!}}}=\f{m\beta}{K^2}\sqrt[n]{n!}=\simeq\f{m\beta}{K^2}\sqrt[n]{\sqrt{2\pi n}\Le(\f ne\Ri)^n}\simeq\f{m\beta}{eK^2}n
\end{align*}
where we approximated the factorial via the Stirling's approximation. From the last identity we get that $r^\free=+\infty$. It then follows that a sufficient condition for a correlation function not to decay as a Gaussian is that its radius of convergence is finite. We can  compare in two regimes of low and high $\beta$ the sequence of the $\alpha_n$ for an ideal gas, for the nonlinear regime and for the linearized potential (harmonic limit). The results are shown in the following figure.
\begin{figure}[H]\label{fig_series}
	\begin{center}
		\includegraphics[scale=0.55]{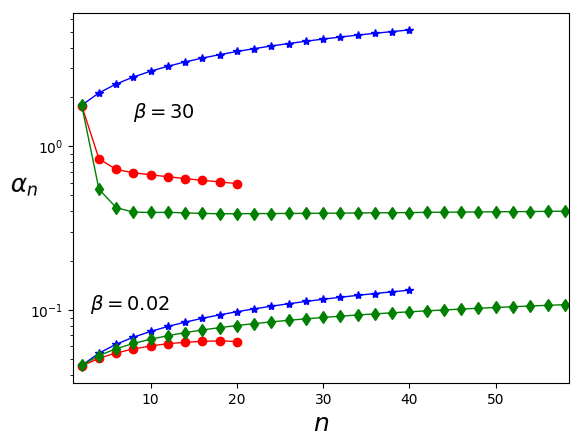}
		\caption{Sequence of $\alpha_n$ for the ideal gas limit (blue stars), for the nonlinear dynamics from the MD simulations (red dots) and for the harmonic regime (green diamonds).}
	\end{center}
\end{figure}
We can see that for both the temperatures the sequence of the $\alpha_n$ for the nonlinear and harmonic dynamics tends to a finite \textit{plateau}; the ideal gas case shows instead a linear increase as expected. As discussed above, the finite asymptotic for the dynamic regimes is sufficient to predict a non-Gaussian relaxation.  
\subsection{Estimate of the high-order kernel coefficients} \label{high_ord_kappas}
In this appendix we derive the estimate eqn.~\ref{high_kap}. \\
We can expand eqn.~\ref{rec_kappa} as 
\begin{align}
\kappa_{2n-2}^I =& \omega^I_{2n} + \sum_{m=1}^{n-1} \chi^{(1)}_{m} \omega^I_{2(n-m)}\omega^I_{2m} + \sum_{m=1}^{n-1}\sum_{m'=1}^{m-1} \chi^{(2)}_{m,m'} \omega^I_{2(n-m)}\omega^I_{2(m-m')}\omega^I_{2m'} + \cdots \nn \\
=& \omega^I_{2n} \left( 1 + \sum_{m=1}^{n-1} \chi^{(1)}_{m} \frac{\omega^I_{2(n-m)}\omega^I_{2m}}{\omega^I_{2n}}+ \sum_{m=1}^{n-1}\sum_{m'=1}^{m-1} \chi^{(2)}_{m,m'} \frac{\omega^I_{2(n-m)}\omega^I_{2(m-m')}\omega^I_{2m'}}{\omega^I_{2n}}  + \cdots \right) \label{exp_kap_n}
\end{align}
with suitable coefficients $\chi^{(j)}_{m_1,\cdots, m_j}$. We can explicitly show that the non-constant contributions in eqn.~\ref{exp_kap_n} vanish via eqn.~\ref{om_i};
starting from the first terms we get:
\begin{align}
\frac{\omega^I_{2(n-m)}\omega^I_{2m}}{\omega^I_{2n}} &=(1-C_{\beta})\frac{(2(n-m))!}{(n-m)!}\frac{(2m)!}{m!}\frac{n!}{(2n)!} \leq 2(1-C_{\beta})\frac{(2(n-1))!}{(n-1)!}\frac{n!}{(2n)!} \label{rat_om} \\
&\leq 2(1-C_{\beta})\frac{n}{2n(2n-1)} \xrightarrow{ n \to \infty } 0 \nn
\end{align}
where in the second line we noticed that $m=1$ corresponds to a maximum of the function
\begin{align}
r_{n,m} \equiv \frac{(2n-2m)!}{(n-m)!}\frac{(2m)!}{m!}
\end{align}
This is shown by evaluating the increment
\begin{align*}
r_{n,m+1} =& \frac{(2n-2m-2)!}{(n-m-1)!}\frac{(2m+2)!}{(m+1)!} 	=\frac{(2n-2m)!}{(2n-2m)(2n-2m-1)}\frac{n-m}{(n-m)!}\f 1{m!(m+1)} (2m)!(2m+1)(2m+2) = \\
=& r_{n,m} \frac{n-m}{(2n-2m)(2n-2m-1)} \frac{(2m+1)(2m+2)}{(m+1)} = r_{n,m}\frac{2m+1}{2n-2m-1} = r_{n,m}f_{n,m}
\end{align*}
where we defined
\begin{equation}
f_{n,m}\equiv \frac{2m+1}{2n-2m-1}
\end{equation}
We can see that the discrete derivative of $f_{n,m}$ as a function of $m$ us always positive:
\begin{align*}
f_{n, m+1}-f_{n,m} = \f{4n}{(1+2m-2n)(3+2m-2n)} \ge \f{4n}{(3+2m-2n)^2} \ge 0
\end{align*}
The denominator of the last identity is always defined for $m,n\in\mathbb N$. It then follows that $f_{n,m}$ is increasing with $m$ and
\begin{equation}
f_{n,m} = 1 \iff  m = \frac{n-1}{2}
\end{equation}
Thus, $r_{n,m+1} \leq r_{n,m}$ if $m\leq (n-1)/2$ and vice-versa. Hence, the maximum values of $r_{n,m}$ are obtained at the boundaries of the admitted values of $m$, i.e. $m=1$ and $m=n-1$, as $r_{n,1}=r_{n, m-1}$.
\subsection{Long-time decay of the memory kernel}\label{tail_ker}
In this section we estimate the long-time tail of $K_j^I(t)$ in Fig.~\ref{kers}, in the high temperature regime. From Fig.~\ref{conv_kap} and Appendix \ref{high_ord_kappas} we know that $\kappa_{2n}^I\simeq \omega_{2n+2}\equiv \kappa_{2n+2}^I$ for $n\gtrsim 40$. We can extend this estimate for any value of $n$ and sum the resulting function, by assuming that for $t\gg 1$ the higher series coefficients matter the most. The first issue that needs to be checked is whether the resulting series is convergent. From the ratio test we get: 
\begin{align*}
\lim_{n\to+\infty} &\llv \f{t^{2n+2}\kappa_{2n+2}^I}{(2n+2)!}\f{(2n)!}{t^{2n}\kappa_{2n}^I}\rrv = t^2\lim_{n\to+\infty} \llv \f{a^{n+2}H_{2n+4}}{(2n+2)!}\f{(2n)!}{a^{n+1}H_{2n+2}}\rrv = at^2\lim_{n\to+\infty} \llv \f{(-2)^{n+2}(2n+3)!!(2n)!}{(2n+2)!(-2)^{n+1}(2n+1)!!} \rrv =\\
&=2at^2 \lim_{n\to+\infty}\llv \f{\f{(2(n+2))!}{2^{n+2}(n+2)!}(2n)!}{(2n+2)!\f{(2(n+1))!}{2^{n+1}(n+1)!}} \rrv = 2at^2 \lim_{n\to+\infty} \f{(2n+4)(2n+3)}{2(2n+2)(2n+1)(n+2)} = at^2\lim_{n\to+\infty}\f 1 n =0
\end{align*}
This shows that the radius of convergence of the series obtained with the approximation $\kappa_{2n}\simeq\kappa_{2n}^I$ is infinite; we can therefore sum the approximation of the tail for any value of $n$: \\
\begin{align}
& K_j(t) \simeq_{t\gg 1} \sum_{n=0}^{+\infty}\f{t^{2n}}{(2n)!} \omega^I_{2n+2} = \sum_{n=0}^{+\infty}\f{t^{2n}}{(2n)!} \Le[(1-C_\beta)a^{n+1}H_{2n+2}+\delta_{n+1,0}C_\beta\Ri] = \nn \\
&=\sum_{n=0}^{+\infty}\f{t^{2n}}{(2n)!}\Le[(1-C_\beta)a^{n+1}(-2)^{n+1}(2n+1)!!\Ri] = (1-C_\beta)\sum_{n=0}^{+\infty}\f{t^{2n}}{(2n)!}a^{n+1}(-2)^{n+1}\f{2^{n+1}\Gamma(n+\f 32)}{\sqrt \pi} = \nn \\
&= (1-C_\beta)\sum_{n=0}^{+\infty}(-1)^{n+1}\f{t^{2n}}{(2n)!}a^{n+1}\f{4^{n+1}}{\sqrt{\pi}}\f{\sqrt{\pi}(2n+2)!}{4^{n+1}(n+1)!}=(1-C_\beta)a\sum_{n=0}^{+\infty}(-1)^{n+1}\Le(at^2\Ri)^n\f{(2n+2)(2n+1)}{(n+1)!}= \nn \\
&=(1-C_\beta)a\sum_{n=0}^{+\infty}(-1)^{n+1}\Le(at^2\Ri)^n\f{(2n+2)(2n+1)}{(n+1)!}
= 2(1-C_\beta)a e^{-at^2}(2at^2-1) \label{ker_long}
\end{align}
which means
\begin{equation}\label{est_ker}
\f{K_j(t)}{K_j(0)} \simeq_{t\gg 1} e^{-at^2}(1-2at^2)
\end{equation}
From the result above we can choose the following ansatz for the long time limit:
\begin{equation*}
K_j(t)\simeq_{t\gg 1} A t^2 e^{-Bt^2}
\end{equation*}
The coefficients $A$ and $B$ can be fixed for each temperature by extending the sum in eqn.~\ref{Knmax} continuously with continuous derivative from a time $t^*=t^*(\beta)$ before its divergence The parameters $A$ and $B$ are then fixed by
\begin{align*}
A &= \f{K^*}{{t^*}^2 \exp\Le(\f{{K'}^*t^* - 2 K^* }{2K^* }\Ri) } \\
B &=  \f{2K^*-{K'}^* t^*}{2K^* {t^*}^2}
\end{align*}
where
\begin{align*}
K^* &\equiv K_{N_{max}}(t^*) \\
{K'}^* &\equiv \sum_{n=0}^{N_{max}} \f{\kappa_{2n}}{(2n-1)!}{t^*}^{2n-1}
\end{align*}
We can additionally check that the series obtained by summing the first neglected terms in the expansion of the kernel eqn.~\ref{exp_kap_n} is convergent. Although this is far from proving the boundedness of the infinitely many orders neglected, it represents a necessary condition. The time-series of the first order correction is given by
\begin{align}
&K_1^I(t)\equiv\sum_{n=0}^{+\infty} \f{t^{2n}}{(2n)!}\sum_{m=1}^n\chi_m^{(1)}\omega_{2(n-m+1)}^I\omega_{2m}^I  
=\sum_{n=0}^{+\infty}\f{t^{2n}}{(2n)!}\sum_{m=1}^n\chi_m^{(1)}\Le[(1-C_\beta)a^{n-m+1}H_{2(n-m+1)}\Ri]\Le[(1-C_\beta)a^m H_{2m}\Ri] = \nn\\
&= - (1-C_\beta)^2\sum_{n=0}^{+\infty} a^{n+1}\f{t^{2n}}{(2n)!}\sum_{m=1}^n (-2)^{n-m+1}\Le(2(n-m)+1\Ri)!!(-2)^m(2m-1)!!=\nn \\
&=-(1-C_\beta)^2\sum_{n=0}^{+\infty}(-2a)^{n+1}\f{t^{2n}}{(2n)!}\sum_{m=1}^n\f{\Le(2(n-m+1)\Ri)!}{2^{n-m+1}(n-m+1)!}\f{(2m)!}{2^m m!} = \nn \\
&= - (1-C_\beta)^2a\sum_{n=0}^{+\infty}\f{\Le(-at^2\Ri)^n}{(2n)!}\sum_{m=1}^n\f{\Le(2(n-m+1)\Ri)!}{(n-m+1)!}\f{(2m)!}{m!}\equiv\sum_{n=0}^{+\infty}a_n \label{series_1ord}
\end{align}
where in the second line we used the property $\chi^{(1)}_m=-1\;\;\forall m$, as it can be explicitly checked from the recursive reconstruction of the relation eqn.~\ref{rec_kappa} till high orders. Following the same argument presented in Appendix \ref{high_ord_kappas}, we can determine
\begin{equation*}
\max_{m\in\Le\{1,n\Ri\}}\f{\Le(2(n-m+1)\Ri)!}{(n-m+1)!}\f{(2m)!}{m!} = \Le.\f{\Le(2(n-m+1)\Ri)!}{(n-m+1)!}\f{(2m)!}{m!}\Ri|_{\substack{m=1\\m=n}} = 2\f{(2n)!}{n!} 
\end{equation*}
We can then check the convergence of the series in eqn.~\ref{series_1ord} via the direct comparison test:
\begin{align*}
\llv a_n \rrv \le \llv -(1-C_\beta)^22a\f{ (-at^2)^n}{(2n)!}n\f{(2n)!}{n!}\rrv\equiv \llv b_n \rrv = \llv -(1-C_\beta)^22a\f{ (-at^2)^n}{(n-1)!}\rrv\equiv \llv b_{n-1} \rrv, \hspace{5mm} n\ge 1
\end{align*}
Via the ratio test we can notice that the series $\sum_{n=0}^{+\infty} b_n$ is convergent:
\begin{align*}
\lim_{n\to+\infty}\llv \f{b_{n+1}}{b_n}\rrv = \lim_{n\to+\infty} \llv -a t^2\rrv\f{(n-1)!}{n!} = 
\llv -a t^2\rrv \lim_{n\to+\infty}\f 1 n =0
\end{align*}
which then proves the convergence of eqn.~\ref{series_1ord}.
\subsection{Management of high-dimensional tensors}\label{map_tensor}
The components of $\mc I_{\mb{ms}}^{(n)}$ can be stored in a one-dimensional pointer in row-major order. For example the indeces of a $3\times 3\times 3$ tensor would be sorted as
\begin{equation*}
\Le\{000,001,002,010,011,012,021,\cdots, 222\Ri\}
\end{equation*}
and by mapping the sequence of the indexes into integer numbers according to 
\begin{equation*}
(i, j, k)\rightarrow i \cdot10^2 + j\cdot 10 +k
\end{equation*}
we get a monotonically increasing sequence. 
In this case, let us define  a general label
\begin{equation*}
\mb v = \{v_0, \cdots, v_{\nind-1}\}
\end{equation*}
of a tensor $\mc A$ of dimension $\mb D =\Le\{D_0, \cdots, D_{\nind-1}\Ri\} \in \mathbb N^{n_{ind}}$ and complex images; $D_j \equiv \car\Le(\Le\{v_j\Ri\}\Ri)$ is the cardinality of the set of the allowed values of the $v_j$. In our case $v_j\in\mathbb N\cup\{0\}$ is a power of a certain coordinate of the system; it follows that $v_j\in\{0,1,\cdots, D_j-1\}$.
Any entry of the tensor $\mc A_{\mb v}\in \mathbb C$ can be mapped as an entry of a one dimensional array via the function
\begin{equation}
\map(\mb v, \mb D, \nind)=v_0\prod_{\beta_0=1}^{\nind-1} D_{\beta_0} + v_1\prod_{\beta_1=2}^{n_{ind}-1} D_{\beta_1} + \cdots v_{\nind-1} = \sum_{\alpha=0}^{n_{ind}-1} v_\alpha\prod_{\beta_\alpha=\alpha+1}^{n_{ind}-1} D_{\beta_\alpha}\label{map_def}
\end{equation}
It is straightforward to determine the map of last index of the tensor $\mb v_{last} \equiv \{D_0-1, D_1-1, \cdots, D_{\nind-1}-1\}$:
\begin{align*}
\map(\mb v_{last}, \mb D, \nind)&= \sum_{\alpha=0}^{\nind-1} (D_\alpha-1)\prod_{\beta_\alpha=\alpha+1}^{n_{ind}-1} D_{\beta_\alpha} =\Le[ (D_0-1)D_1 D_2\cdots D_{\nind-1}\Ri] +\\ 
&+\Le[(D_1-1)D_2\cdots D_{\nind-1}\Ri]+\cdots+ D_{\nind-1}-1 =\prod_{\gamma=0}^{\nind-1} D_\gamma -1 \equiv M_{\mb D}
\end{align*}
The identity follows as the sum is telescopic. \\ \\
In the implementation of the recursion relations in Section \ref{constr} controlling $(i\mc L)^n e^{iKr_j}\rightarrow (i\mc L)^{n+1}e^{iK r_j}$ we are interested in integer increments (let us say by $l\in\mathbb N$) of a generic $i$-th 'column' of the coefficients tensor: $\mc I^{(n)}_{\mb{ms}}\rightarrow I^{(n)}_{\mb{ms}+l\mb{\hat e_i}}$. By imposing $v_i\equiv l \in\{0, \cdots, D_i-1\}$, it directly follows from eqn.~\ref{map_def}:
\begin{align*}
\map(\{v_0,\cdots, v_{i-1},l,v_{i+1},\cdots, v_{\nind-1} \}, \mb D, \nind) = \sum_{\substack{\alpha=0\\\alpha\neq i}}^{\nind-1} v_\alpha\prod_{\beta_\alpha=\alpha+1}^{n_{ind}-1} D_{\beta_\alpha} + l \prod_{\beta_i=i+1}^{n_{ind}-1} D_{\beta_{i}}
\end{align*} 
It is moreover possible to determine the inverse map that, given an entry of a one-dimensional mapping and the list of the related dimensions, returns the multi-index associated to that component. For this we can proceed recursively, from the extraction of the last index $v_{\nind-1}$ backwards. From eqn.~\ref{map_def} we get: 
\begin{align}
\nmap\equiv\map(\mb v, \mb D, \nind) &= \Le(\sum_{\alpha=0}^{\nind-2}v_{\alpha}\prod_{\beta_\alpha=\alpha+1}^{\nind-1}D_{\beta_\alpha}\Ri)+v_{\nind-1} = \Le(\sum_{\alpha=0}^{\nind-2}v_{\alpha}\prod_{\beta_\alpha=\alpha+1}^{\nind-2}D_{\beta_\alpha}\Ri)D_{\nind-1}+v_{\nind-1}\label{map1}\\
v_{\nind-1}&=\nmap \mod D_{\nind-1} \label{map2}
\end{align}
Eqn.~\ref{map2} follows from eqn.~\ref{map1} being $v_{\nind-1}<D_{\nind-1}$.
We can proceed with the extraction of the second last component $v_{\nind-2}$ via the knowledge of $v_{\nind-1}$:
\begin{align*}
A_{n_{ind}-1}\equiv\f{\nmap-v_{\nind-1}}{D_{\nind-1}} &= \Le(\sum_{\alpha=0}^{\nind-3} v_\alpha \prod_{\beta_\alpha=\alpha+1}^{\nind-3} D_{\beta_\alpha}\Ri) D_{\nind-2} + v_{\nind-2} \\
v_{\nind-2} & = A_{n_{ind}-1} \mod D_{\nind-2}
\end{align*}
A recursion relation can then be established $\forall \; \;i\in\{0,\cdots, \nind-1\}$:
\begin{align*}
v_{\nind-i-1}&=A_{\nind-i}\mod D_{\nind-i-1} \\
A_{\nind-i-1} &= \f{A_{\nind-i} - v_{\nind-i-1}}{D_{\nind-i-1}}
\end{align*}
with initial condition $A_{\nind}=\nmap$. 

\end{appendices}

\bigskip
\section*{Acknowledgments}
We thank T.~Voigtmann, T.~Franosch and A.~Zippelius for useful discussions. Computer simulations presented in this paper were carried out using the bwForCluster NEMO high-performance computing facility.

\bibliography{biblio}

\begin{thebibliography}{10}

\bibitem{dna}
M.~Peyrard and J.~Farago, ``{Nonlinear localization in thermalized lattices:
  application to DNA},'' {\em Physica A: Statistical Mechanics and its
  Applications}, vol.~288, no.~1, pp.~199 -- 217, 2000.

\bibitem{poly}
A.~Henry and G.~Chen, ``{Anomalous heat conduction in polyethylene chains:
  Theory and molecular dynamics simulations},'' {\em Phys. Rev. B}, vol.~79,
  p.~144305, Apr 2009.

\bibitem{spohn_growth}
H.~Spohn, ``{Exact solutions for KPZ-type growth processes, random matrices,
  and equilibrium shapes of crystals},'' {\em Physica A: Statistical Mechanics
  and its Applications}, vol.~369, no.~1, pp.~71 -- 99, 2006.

\bibitem{gavallotti}
G.~Gallavotti, {\em The Fermi-Pasta-Ulam Problem: A Status Report}.
\newblock Berlin, Heidelberg: Springer, 2008.

\bibitem{Bambusi2015}
D.~Bambusi, A.~Carati, A.~Maiocchi, and A.~Maspero, {\em Some Analytic Results
  on the FPU Paradox}, pp.~235--254.
\newblock New York, NY: Springer New York, 2015.

\bibitem{fermi_pasta_ulam1955}
E.~Fermi, J.~Pasta, and S.~Ulam, ``Studies of nonlinear problems {I}, {L}os
  {A}lamos {R}eport {LA} 1940, 1955,'' 1974.

\bibitem{dauxois}
T.~Dauxois, ``{Fermi, Pasta, Ulam, and a mysterious lady},'' {\em Physics
  Today}, vol.~61, no.~1, pp.~55--57, 2008.

\bibitem{fermipoincare1}
E.~Fermi, ``{Beweiss das ein mechanisches Normalsystem im allgemeinen
  quasi-ergodisch ist},'' {\em Phys. Z.}, vol.~24, p.~261, 1923.

\bibitem{Rink2006}
B.~Rink, ``{Proof of Nishida's Conjecture on Anharmonic Lattices},'' {\em
  Communications in Mathematical Physics}, vol.~261, no.~3, pp.~613--627, 2006.

\bibitem{Friesecke2014}
G.~Friesecke and A.~Mikikits-Leitner, ``{Cnoidal Waves on Fermi–Pasta–Ulam
  Lattices},'' vol.~27, 01 2014.

\bibitem{hajnal}
D.~Hajnal and R.~Schilling, ``Delocalization-localization transition due to
  anharmonicity,'' {\em Phys. Rev. Lett.}, vol.~101, p.~124101, Sep 2008.

\bibitem{caratieq}
A.~Carati and L.~Galgani, ``{Metastability in specific-heat measurements:
  Simulations with the FPU model},'' {\em EPL (Europhysics Letters)}, vol.~75,
  no.~4, p.~528, 2006.

\bibitem{maiocchi2013}
A.~Maiocchi, D.~Bambusi, and A.~Carati, ``An averaging theorem for fpu in the
  thermodynamic limit,'' {\em Journal of Statistical Physics}, vol.~155,
  pp.~300--322, Apr 2014.

\bibitem{maiocchi}
A.~M. Maiocchi, A.~Carati, and A.~Giorgilli, ``A series expansion for the time
  autocorrelation of dynamical variables,'' {\em Journal of Statistical
  Physics}, vol.~148, pp.~1054--1071, Sep 2012.

\bibitem{carati2015}
A.~Carati, A.~Maiocchi, L.~Galgani, and G.~Amati, ``{The Fermi--Pasta--Ulam
  System as a Model for Glasses},'' {\em Mathematical Physics, Analysis and
  Geometry}, vol.~18, p.~31, Nov 2015.

\bibitem{mori}
H.~Mori, ``{Transport, Collective Motion, and Brownian Motion},'' {\em Progress
  of Theoretical Physics}, vol.~33, no.~3, pp.~423--455, 1965.

\bibitem{zwanzig}
R.~Zwanzig, ``{Memory Effects in Irreversible Thermodynamics},'' {\em Phys.
  Rev.}, vol.~124, pp.~983--992, Nov 1961.

\bibitem{grabert}
H.~Grabert, {\em Projection operator techniques in nonequilibrium statistical
  mechanics}.
\newblock Berlin: Springer, 1982.

\bibitem{wierling}
A.~Wierling, ``Dynamic structure factor of linear harmonic chain -- a
  recurrence relation approach,'' {\em The European Physical Journal B},
  vol.~85, Jan 2012.

\bibitem{hugues}
H.~Meyer, T.~Voigtmann, and T.~Schilling, ``{On the non-stationary generalized
  Langevin equation},'' {\em The Journal of chemical physics}, vol.~147,
  p.~214110, Aug 2017.

\bibitem{allentid}
M.~P. Allen and D.~J. Tildesley, {\em Computer Simulation of Liquids}.
\newblock New York, NY, USA: Clarendon Press, 1989.

\bibitem{reichman}
D.~R. Reichman and P.~Charbonneau, ``{Mode-coupling theory},'' {\em Journal of
  Statistical Mechanics: Theory and Experiment}, vol.~2005, no.~05, p.~P05013,
  2005.

\bibitem{hansenmd}
J.-P. Hansen and I.~R. McDonald, eds., {\em Theory of Simple Liquids (Fourth
  Edition)}.
\newblock Oxford: Academic Press, 2013.

\bibitem{stretched_exp}
B.~Jean‐Philippe, {\em Anomalous Relaxation in Complex Systems: From
  Stretched to Compressed Exponentials}, ch.~11, pp.~327--345.
\newblock Wiley-Blackwell, 2008.

\bibitem{yoshida}
T.~Yoshida, K.~Shobu, and H.~Mori, ``Dynamic properties of one-dimensional
  harmonic liquids. idensity correlation and transport coefficients,'' {\em
  Progress of Theoretical Physics}, vol.~66, no.~3, pp.~759--771, 1981.

\bibitem{shobu}
K.~Shobu, T.~Yoshida, and H.~Mori, ``Dynamic properties of one-dimensional
  harmonic liquids. ii— energy density correlation and heat transport,'' {\em
  Progress of Theoretical Physics}, vol.~66, no.~4, pp.~1160--1168, 1981.

\bibitem{radons}
G.~Radons, J.~Keller, and T.~Geisel, ``Dynamical structure factor of a
  one-dimensional harmonic liquid: Comparison of different approximation
  methods,'' {\em Zeitschrift f{\"u}r Physik B Condensed Matter}, vol.~50,
  pp.~289--296, Dec 1983.

\end{thebibliography}
\bibliographystyle{ieeetr}
\end{document}